\documentclass[lettersize,10pt]{article}
\usepackage{graphicx}
\usepackage{amsmath}
\usepackage{amsfonts}
\usepackage{bm}
\usepackage[margin=.75in]{geometry}

\newcommand{\dd}{\text{d}}

\newcommand{\e}{\text{e}}
\newcommand{\vv}{{\bf v}}
\newcommand{\zz}{{\bf 0}}
\newcommand{\shat}{\hat{\sigma}}
\newcommand{\rhat}{\boldsymbol{\hat{r}}}
\newcommand{\that}{\boldsymbol{\hat{\theta}}}

\title{Assessment of the Effects of Azimuthal Mode Number Perturbations upon the Implosion Processes of Fluids in Cylinders}

\author{Michael Lindstrom \thanks{Mathematics Department, University of California, Los Angeles
{\tt mikel@math.ucla.edu} (corresponding author)} 
}

\begin{document}

\maketitle

\begin{abstract}

Fluid instabilities arise in a variety of contexts and are often unwanted results of engineering imperfections. In one particular model for a magnetized target fusion reactor, a pressure wave is propagated in a cylindrical annulus comprised of a dense fluid before impinging upon a plasma and imploding it. Part of the success of the apparatus is a function of how axially-symmetric the final pressure pulse is upon impacting the plasma. We study a simple model for the implosion of the system to study how imperfections in the pressure imparted on the outer circumference grow due to geometric focusing. Our methodology entails linearizing the compressible Euler equations for mass and momentum conservation about a cylindrically symmetric problem and analyzing the perturbed profiles at different mode numbers. The linearized system gives rise to singular shocks and through analyzing the perturbation profiles at various times, we infer that high mode numbers are dampened through the propagation. We also study the Linear Klein-Gordon equation in the context of stability of linear cylindrical wave formation whereby highly oscillatory, bounded behaviour is observed in a far field solution. 

\end{abstract}

{\bf keywords:} 
magnetized target fusion, cylindrical implosion, linear perturbations of conservation laws, singular shocks, asymptotics, linear Klein-Gordon equation, compressible Euler equations


\section{Introduction}

Implosion is the process by which an object is destroyed by being forced to close in on itself. Implosions have many scientific applications such as 
 how gas bubbles, sites for chemical reactions, are eliminated inside of fluids \cite{implodeGas}, the formation and collapse of stars \cite{implodeStar}, or even the demolition of buildings \cite{demo}. 
 There has been mathematical interest in their study such as in 
 finding exact similarity solutions to the porous media equations subject to fixed spatial pressure functions \cite{similarity}, and in studying the stability of similarity solutions in the presence of perturbations \cite{pertSim}. From an engineering perspective, implosions are now being studied in applications to nuclear fusion energy \cite{implodeFus}; in particular with this paper, we will be interested in understanding the implosion relevant to the Magnetized Target Fusion (MTF) reactor being designed by the Canadian fusion energy research company General Fusion \cite{GF}.

The essence of the MTF reactor is to implode a giant sphere of molten metal (lead-lithium), at whose centre is a plasma fuel (deuterium and tritium) held in place by a magnetic field. Through a focusing effect, a pressure wave inside the molten metal moves radially inward and increases in strength until it impacts the plasma, imparting an immense amount of energy that causes the plasma to compress and fuse. The fusion process then releases energy. This setup has been studied from a variety of angles including numerical simulations of models \cite{GF} \cite{FV-Num}, via formal asymptotic analysis \cite{asy}, and more recently in studying the Richtmyer-Meshkov (RM) instability between an imploding molten metal and a plasma fuel in the context of azimuthal asymmetries \cite{GF-RM}. An RM instability is formed between two fluids joined at an interface undergoing an acceleration whereby fingering or mixing-type behaviour may be observed \cite{RM} \cite{RMExp}. Such phenomena are typically modelled as the interaction of viscous fluids and popular methods of analysis entail linear perturbations with different mode numbers or numerical simulations. Another related, although different fluid instability is the Taylor-Rayleigh instability that emerges when a fluid pushes a denser fluid \cite{RT}. This instability can pose problems for applications such as Inertial Confinement Fusion \cite{ICF}. In the context of MTF, the mixing of the fuel with the molten metal could greatly reduce the efficiency of the apparatus. Our problem will study a single inviscid fluid with a shock wave, which could in some senses be interpreted as two separate fluids as the density is different on either side of the shock. Similar to \cite{GF-RM}, we will also be concerned with the effects of asymmetries, but our approach will use a linearization of an axially symmetric implosion, which gives rise to hyperbolic equations in one spatial dimension with singular shocks, instead of numerically solving a two-dimensional system describing the plasma and molten metal interaction directly.

Hyperbolic partial differential equations may not admit a strong solution and at times a weak, discontinuous solution is obtained. Valuable background into the theory of partial differential equations can be found in \cite{Evans}. There are some equations and scenarios where a singular shock emerges: a solution that includes a measure-valued function, localized to a single point or surface, moving through space. The theoretical properties of singular shock solutions have been studied quite extensively for a variety of 
 equations such as with the inviscid Burgers equation \cite{SSTheoryBurg1} and with the equations of geometric optics where the singular shocks are also observed numerically \cite{GeoOpt}. Despite the theoretical interest, such solutions arise naturally in certain fluid flow problems, such as those describing the high concentration limits of particles suspended in viscous fluids \cite{Andrea} and in the case of this paper in linearizing conservation laws. 
  There have also been sophisticated numerical methods developed for solving problems with perturbed shock fronts in fewer iterations than what standard first-order methods may yield \cite{zua}, and even within the community of those studying elliptic and parabolic partial differential equations delta functions are a topic of study \cite{delPara}.

\begin{figure}
 \begin{center}
    \includegraphics[width=3.5in]{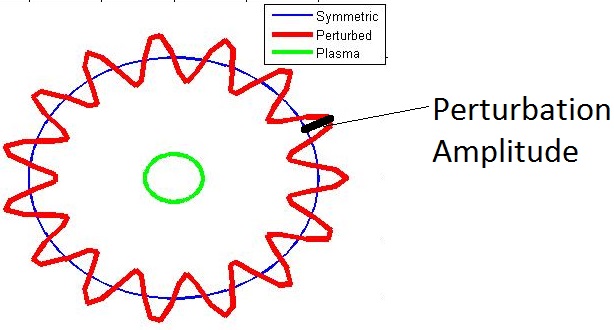}
 \end{center}
\caption{A plasma is found at the centre of the apparatus. Through the impulse of the pistons, a shock wave is generated on the outside and moves radially inward. The blue curve represents the implosion under perfect symmetry, where a shock front would be found, and the red curve represents the shock front when the implosion is not perfectly symmetric.}
\label{fig:imp}
\end{figure}

The work of this paper studies implosions in the context of MTF. In this work we will linearize equations modelling an imploding cylinder of molten metal about an axially symmetric implosion. The perturbations will be carried out over each azimuthal mode number. By studying the strengths of the singular shocks of the perturbed system, i.e., the coefficient multiplying the moving delta function sources, we are able to assess how sensitive the implosion front is to the different mode numbers as seen in figure \ref{fig:imp}. The larger the strength of the delta function, the larger the peak deviation in shock fronts between the symmetric and asymmetric implosion scenarios for a given perturbation size. Thus, we are able to gain insights into the two-dimensional system by solving a one-dimensional problem with singular shocks: a two-by-two nonlinear system modelling the base density and radial momentum density and the three-by-three linear system for the perturbations in the density, radial momentum density, and momentum density in the angular direction.

The work has a number of novel elements. To this author's knowledge, it is the first study of a cylindrical implosion done by linearization in the azimuthal mode number perturbation amplitudes. Through the ensuing equations, singular shocks emerge as solutions to nonhomogeneous conservation laws. Also, through our analysis, we will asymptotically evaluate a Bateman-like integral that does not have a closed form.

The paper is organized as follows: section \ref{MM} provides the motivation for our study and derives the model; following this, we will study delta function solutions in the context of a nonhomogeneous inviscid Burgers' equation and their physical meaning in section \ref{Burg}. While the Burgers' equation has been well studied, because we wish to carefully interpret the physical significance of the delta function solutions and to validate our numerical procedure, the Burgers' equation serves a useful benchmark. We will then study an asymptotic toy model of the implosion to learn some qualitative behaviour of an implosion at high azimuthal mode numbers in section \ref{Asy}. The equation that we ultimately need to solve in section \ref{Asy} is the linear Klein-Gordon equation in one dimension on the half-line with a step function boundary condition. The Klein-Gordon equation often arises in problems pertaining to quantum mechanics \cite{LKG}. Following this work, section \ref{Imp} presents the numerical results obtained in solving the cylindrical model, including discussing their significance, and we conclude our work in section \ref{Con}.

\section{Motivation and Model}

\label{MM}

The design of the MTF reactor proposed by General Fusion \cite{GF} involves a sphere of molten metal of radius $1.5$ m rotating such that an empty cylindrical cavity of radius $0.2$ m is formed along the axis of rotation.  On the order of 100 pistons, timed to hit precisely at the same time, aimed towards the sphere's centre, are driven into the outer wall of the molten metal cavity at a high speed delivering an impact pressure of 2 GPa over a time scale of around 45 $\mu$s. This sends a pressure wave through the metal, increasing in strength due to a radial focusing effect. A plasma is fired along the axis of the cylinder and held in at the centre by magnetic fields so that it receives the focused energy of the impulse. It then undergoes a rapid compression and fusion takes place. The focus of this present work is to gain understanding into the energy transfer through the metal cavity and we will not be concerned with the plasma interaction. Specifically, we wish to study the effects of imperfect symmetry in the implosion: while an ideal setting involves a perfect spherically symmetric or cylindrically symmetric implosion, in reality, with a finite number of pistons, this is not achievable.

For this study, we choose lead as the metal, and we will make some simplifying assumptions. We assume a linear equation of state for lead with \begin{equation} P = P_{\text{atm}} + c_s^2 (\rho - \varrho_0) \label{Peos} \end{equation} where $P$ is the pressure, $P_{\text{atm}}$ is atmospheric pressure, $c_s$ is the sound speed, $\rho$ is the mass density per unit volume, and $\varrho_0$ is the density of the lead at atmospheric pressure. We obtain our data from \cite{Roth}. In reality, the equation of state for lead is nonlinear, however, it has been shown that a linear model, even in these extreme pressure regimes, is a reasonable approximation \cite{phd}. With $c_s = 2090$ m/s, the energy transfer through the metal is rapid and we will neglect the rotation of the molten metal. This will also allow us to link our work with \cite{GF-RM} where rotation is also not present. We use the compressible Euler equations for mass and momentum conservation \cite{Fluid} \begin{align} \rho_t + \nabla \cdot \boldsymbol\mu &= 0 \label{basicmass} \\  \boldsymbol\mu_t + \nabla P + \nabla \cdot (\frac{\boldsymbol\mu}{\rho} \otimes \boldsymbol\mu) &= \zz \label{mombasic} \end{align} where $\rho$ is the mass density and $\boldsymbol\mu$ is the momentum density: the product of $\rho$ and the local fluid velocity $\vv.$

The device design is somewhere between a two- and three-dimensional system in that there are both cylindrical and spherical symmetries present. In order to apply a linearization and suitably diagonalize the equations so that there is only one spatial dimension and time, we adopt a cylindrical model. Our problem will also be independent of the height in the $z-$directions so we will use polar coordinates. We consider the annular region $r_m < r < r_M$, $r_m = 0.2$ m, $r_M = 1.5$ m, filled with molten lead. Ideally, at a time $t=0$ s, pistons impact the outer wall of the cavity imparting a uniform pressure at $r = r_M$ given by $P(r_M,\theta,t) = P_{\text{atm}} + (P_M - P_{\text{atm}}) \e^{-t^2/t_0^2}$ where $P_M$ is the $2$ GPa impact pressure and the Gaussian decay with time scale $t_0 = 45 \mu$s models the piston impulse decay rate. With $N_s \approx 100$ pistons impacting a sphere, each piston takes up a solid angle of $\Omega \approx \frac{4 \pi}{N_s}$ steradians, which would be subtended by a cone with vertex at the sphere centre having angle $\theta$ radians where $\Omega = 2 \pi (1 - \cos(\theta/2)).$ For large $N$, we have that $$\theta \approx \frac{4}{\sqrt{N_s}}.$$ This suggests that if we wish to model the pistons on a cylinder, we can consider on the order of \begin{equation} N_c = \frac{2 \pi}{\theta} = \frac{\pi \sqrt{N_s}}{2} \label{Nc} \end{equation} pistons. With $N_s \approx 100$, then $N_c \approx 15$. If each piston can be ascribed to an angle $\Delta = 2 \pi/N_c$ and each piston impacts that piece of the metal cavity in the radial direction spanning a fraction of the angle $0 \leq F \leq 1$ then an improved estimate for the form of the pressure impulse is $$P(r_M, \theta, t) = P_{\text{atm}} + \sum_{j=0}^{N_c-1} \chi_{[(j-F/2)\Delta, (j+F/2) \Delta]} (P_M - P_{\text{atm}}) \text{e}^{-t^2/t_0^2}$$ where $\chi_I$ denotes the characteristic function on $I$. A more accurate model could include the non-radial impact velocity. Through Fourier analysis, we can write \begin{align*} \sum_{j=0}^{N_c-1} \chi_{[(j-F/2)\Delta, (j+F/2) \Delta]} &= F + \sum_{k = 1}^{\infty} \frac{2}{\pi k} \sin( k \pi F) \cos(k N_c \theta) \\ &= F + \sum_{k=1}^{\infty} \frac{2}{\pi k} \sin(k \pi F) \Re( \e^{i k N_c \theta} ) \end{align*} where the series written will repeat over all of $\mathbb{R}$, but matches appropriately on $[-\pi, \pi].$ This is consistent with physical intuition: if $F = 0$ then the pistons do nothing and each summand is zero; if $F = 1$ then the pulse is axially symmetric and all the terms in the summand are again zero with the total sum of $F = 1.$ Thus, the azimuthal mode numbers that perturb the system from a perfect cylindrical collapse are $m = k N_c$ with amplitude $F$ for $m=0$ and $\frac{2 \sin(k \pi F)}{\pi k}$ for $m>0$. For fixed $k$, these amplitudes can be made $O(1-F)$ if $F$ is close to $1$; and these amplitudes tend to zero with $1/k$ for any $F$.

Initially we assume that the system is at rest at atmospheric pressure. The pressure at the outer boundary is prescribed by the impulse. We remark that the outer boundary of the molten lead does move, but the movement is asymptotically negligible \cite{asy} and therefore we assume the outer boundary is stationary. For the inner boundary conditions, we will choose that the density and velocity at the inner boundary $r = r_m$ are constant in time, but these boundary conditions are irrelevant as we will be stopping the simulations before the point of impact. A table summarizing the physical parameters in the model is listed in table \ref{tab:physical}.

\begin{table}[h]
\begin{center}
\begin{tabular} {c | c }
Parameter & Value \\
\hline
Outer radius of molten metal $r_M$ & 1.5 m \\
Inner radius of molten metal $r_m$ & 0.2 m \\
Sound speed $c_s$ & 2090 m/s \\
Density of lead at atmospheric pressure $\varrho_0$ & 11340 kg/m$^3$ \\
Peak pressure $P_M$ & 2 GPa \\
Pressure time scale $t_0$ & 45 $\mu$s \\
Atmospheric pressure $P_{\text{atm}}$ & 101325 Pa \\
Number of pistons on cylinder $N_c$ & 15
\end{tabular} \caption{Physical parameters for the system.} \label{tab:physical}
\end{center}

\end{table}

To understand the asymmetry in each mode number, we perform a linear perturbation analysis of a cylindrically symmetric system whereby $P(r_M,\theta,t) = P_{\text{atm}} + P_M \e^{-t^2/t_0^2} (1 + \eta \exp(im \theta))$ for a linearization parameter $\eta$ and with a mode number $m$. We use the tensor identities for polar coordinates \begin{align*} \nabla f &= \partial_r f \rhat + \frac{1}{r} \partial_\theta f \that \\ {\bf \nabla} \cdot (f_1 \rhat + f_2 \that) &= \partial_r f_1 + \frac{1}{r} (\partial_\theta f_2 + f_1) \\ \nabla \cdot (f_{11} \rhat \otimes \rhat + f_{12} \rhat \otimes \that + f_{21} \that \otimes \rhat + f_{22} \that \otimes \that) &= \left[ \partial_r f_{11} + \frac{1}{r} (\partial_\theta f_{21} + f_{11} - f_{22}) \right] \rhat \\ &+ \left[ \partial_r f_{12} + \frac{1}{r} (\partial_\theta f_{22} + f_{12} + f_{21}) \right] \that \end{align*} such that with $\boldsymbol\mu = \mu_1 \rhat + \mu_2 \that$, equations (\ref{basicmass}) and (\ref{mombasic}) yield \begin{align} \rho_t + \mu_{1,r} + \frac{1}{r} (\mu_{2,\theta} + \mu_1) &= 0 \label{mass1} \\
 \mu_{1,t} + P_r + (\frac{\mu_1^2}{\rho})_r + \frac{1}{r} \left[ (\frac{\mu_1 \mu_2}{\rho})_\theta + \frac{\mu_1^2 - \mu_2^2}{\rho}  \right] &= 0 \label{mom1} \\
\mu_{2,t} + (\frac{\mu_1 \mu_2}{\rho})_r + \frac{1}{r} \left[ P_\theta + \frac{\mu_2^2 + 2 \mu_1 \mu_2}{\rho} \right] &= 0 \label{mom2} \end{align}

\noindent We formulate a linearized hyperbolic system about a radially symmetric base state with $\rho = \rho_0 (r,t) + \eta \e^{i m \theta} \bar{\rho} (r, t) + ...$, $\mu_1 = \mu_0 (r,t) + \eta \e ^{i m \theta} s (r, t) + ...,$ and $\mu_2 = \eta \e^{i m \theta} \psi (r, t) + ...$. Note that $\mu_2$ starts at $O(\eta)$ as there should be no velocity in the angular direction for a symmetric collapse. 
After nondimensionalizing and assuming there are initially equilibrium conditions, with the only disturbance occurring at the outer boundary
, the system we solve within $a < r < 1$, $t>0$ is:
\begin{align} \rho_{0,t} + \mu_{0,r} + \frac{1}{r} \mu_0 &= 0 \label{rho0} \\
\mu_{0,t} + (c^2 \rho_0 + \mu_0^2/\rho_0)_r + \frac{1}{r} \mu_0^2/\rho_0 &= 0 \label{m0} \\
 \bar{\rho}_{t} + s_r + \frac{1}{r} s + \frac{im}{r} \psi &= 0 \label{rhobar} \\
s_t + (c^2 \bar{\rho} + 2 \mu_0 s/\rho_0 - \mu_0^2 \bar{\rho}/\rho_0^2)_r + \frac{1}{r} (2 \mu_0 s / \rho_0 + \mu_0^2 \bar{\rho} / \rho_0^2) + \frac{im}{r} \mu_0 \psi / \rho_0 &= 0 \label{sbar} \\
\psi_t + (\mu_0 \psi/\rho_0)_r + \frac{2}{r} \mu_0 \psi / \rho_0 + \frac{im}{r} c^2 \bar{\rho} &= 0 \label{wbar}
\end{align}
subject to
\begin{align}
\mu_0 (r,0) = \bar{\rho}(r,0) = s(r,0) = \psi (r,0) &= 0 \label{zeroinits} \\
\rho_0(r,0) &= 1 \\
 \rho_0(a,t) &= 1 \label{rho0left} \\
\mu_0(a,t) = \bar{\rho}(a,t) = s(a,t) = \psi(a,t) &= 0 \label{lefts} \\
\rho_0(1,t) &= (\e^{-t^2/\tau^2} - \delta)/c^2 + 1 \label{rhorts} \\
\bar{\rho}(1,t) &= \e^{-t^2/\tau^2} \label{rights} 
 \end{align} The boundary conditions for $\mu_0$, $s$, and $\psi$ at $r=1$ are determined dynamically from the system, which we discuss more fully in section \ref{Imp}. Note that all these variables and parameters are now dimensionless: density has been nondimensionalized by $\varrho_0$, pressure by $P_M$, length by $r_M$, time by $r_M \sqrt{\varrho_0/P_M}$, and velocity by $\sqrt{P_M/\varrho_0}.$ The values of the dimensionless constants can be found in table \ref{tab:noncon}.

\begin{table}[h]
\begin{center}
\begin{tabular} {c | c }
Parameter & Value \\
\hline
Dimensionless sound speed $c$ & 4.9622 \\
Dimensionless decay time scale $\tau$ & 0.0126 \\ 
Dimensionless inner radius of molten metal $a$ & 0.1333 \\
Dimensionless atmospheric pressure $\delta$ & 0.00005 \\
\end{tabular} \caption{Dimensionless constants.} \label{tab:noncon}
\end{center}

\end{table}

\section{Significance of Delta-Function Solutions}

\label{Burg}

The study presented here induces small azimuthal perturbations to a cylindrically symmetric implosion problem governed by the Euler equations for fluids. The Euler equations are hyperbolic in nature and weak, shock-like solutions are a common phenomena. As a result, the notion of a linearization needs to be treated with care. Fortunately, many details of linearizing hyperbolic problems have been presented in the literature. 
Through such analysis, the linearized problems often require the use of generalized tangent vectors  \cite{Gateaux} \cite{Var} , and in our numerical setting, distributions distributions whereby there may be delta-functions appearing as a natural part of the solution. 
The purpose of this section is to understand physically how such solutions can arise and to gain some physical insight into their significance, along with gaining confidence in the validity of a standard first order method for such a problem. We are not attempting a rigorous formalism or treatment of these solutions for which we would refer the interested reader literature such as \cite{Gateaux} giving many formal properties of perturbed hyperbolic equations and \cite{anotherTheory} that derives singular shocks of fluids as the limit of viscous systems. Indeed, there are open questions in some of the techniques that we employ. 

We begin by considering the analytic solution to a perturbed nonhomogeneous inviscid Burgers' equation.

\subsection{Perturbed Burgers' Equation - Analytic Approach}

\subsubsection{Generalized Tangent Vector}

We consider the Burgers' equation for a scalar $u$ having flux $f(u) = \frac{1}{2} u^2$ and with source term $u^2$ such that

 \begin{equation} u_t + (\frac{1}{2}u^2)_x = u^2. \label{GM} \end{equation}

We wish to solve (\ref{GM}) subject to the initial conditions \begin{equation} u(x,0) = u^0 (x) = \begin{cases} 1, \quad x < 0 \\ 0, \quad x \geq 0 \end{cases} \label{IC0} \end{equation} and \begin{equation} u(x,0) = u^p (x) = \begin{cases} 1 - \eta, \quad x < 0 \\ \eta, \quad x \geq 0 \end{cases} \label{ICp} \end{equation} where $0 < \eta \ll 1$. The two sets of initial conditions are a base set and a perturbed set respectively. We make the remark that due to the discontinuities, we are seeking weak solutions because the strong solutions do not exist.

We can linearize the flux in (\ref{GM}) to read $$u_t + u u_x = u^2$$ so that from the method of characteristics, we know that in (\ref{GM}), $\frac{\dd u}{\dd t} = u^2$ along $\frac{\dd x}{\dd t} = u.$ Thus, from our initializations, the solution should be a piecewise constant and the constants grow in time with $u = \frac{u_0}{1 - u_0 t}$ ($u_0$ is the initial value along a characteristic curve). We note the solution blows up in finite time at $t=1$, so we will only ever consider times less than $t=1.$

With (\ref{IC0}), the characteristics move with velocity $\frac{1}{1-t}$ where $u = \frac{1}{1-t}$ and with velocity $0$ where $u =0$ and thus there is a shock.  The Rankine-Hugoniot \cite{RK} conditions impose that the shock has a velocity $\frac{\dd x_s^0}{\dd t} = [\frac{1}{2} u^2]/[u] = \frac{1}{2(1-t)}$ where $[q]$ denotes the difference of the right- and left-limits of the quantity $q$ across the shock and $x_s^0$ is the shock position. With $x_s^0(0) = 0$, $x_s^0(t) = \frac{-1}{2} \log(1-t)$ and we can find the solution to (\ref{GM}) and (\ref{IC0}) for all time and obtain: 
\begin{equation} u(x,t) = u^{0,s}(x,t) = \begin{cases} \frac{1}{1-t} , \quad x < x_s^0(t) \\ 0, \quad x \geq x_s^0(t). \end{cases} \label{sol0} \end{equation} Similarly, solve the same equation subject to (\ref{ICp}) to obtain \begin{equation} u(x,t) = u^{p,s} (x,t) = \begin{cases} \frac{1-\eta}{1 - (1-\eta)t}, \quad x < x_s^p(t) \\  \frac{\eta}{1 - \eta t}, \quad x \geq  x_s^p(t) \end{cases} \label{solp} \end{equation} with $x_s^p(t) = \frac{-1}{2} (\log(1 - (1-\eta)t) + \log(1-\eta t)).$

In looking at the difference of the solutions (\ref{sol0}) and (\ref{solp}), we observe that $$u^{p,s} - u^{0,s} = 
  \begin{cases} \frac{1-\eta}{1 - (1-\eta)t} - \frac{1}{1-t}, \quad x < x_s^p(t) \\ \frac{\eta}{1 - \eta t} - \frac{1}{1-t} , \quad x_s^p(t) \leq x  < x_s^0(t) \\ \frac{\eta}{1 - \eta t} , \quad x > x_s^0(t) \end{cases}
$$ such that no matter how small $\eta$ may become, there will always be an $O(1)$ difference between the two solutions on some set $(x_s^p(t), x_s^0(t))$, which vanishes to a point as $\eta \rightarrow 0$ and away from this region, there is an $O(\eta)$ difference between the solutions. 

This observation leads to considering a generalized tangent vector, a pair $(\hat{u}, \hat{x}) \in L^1 \times \mathbb{R}$, such that the solution for small $\eta$ can be approximated by starting with $u^{0,s}$, adding $\eta \hat{u}$, and shifting by $\eta \hat{x}$ to account for the small change in shock speeds between $u^{p,s}$ and $u^{0,s}$ \cite{Var}. We identify $\hat{u} = \lim_{\eta \rightarrow 0} \frac{u^{p,s} - u^{0,s}}{\eta}$ with \begin{equation} \hat{u} = \begin{cases} \frac{-1}{(1-t)^2}, \quad x < x_s^0(t) \\ 1, \quad x > x_s^0(t) \end{cases} \label{funVec} \end{equation} and \begin{equation} \hat{x} = \lim_{\eta \rightarrow 0} \frac{x_s^0 - x_s^p}{\eta} = \frac{-t^2}{2(1-t)}. \label{shockVec} \end{equation}


We take the generalized tangent vector notion a step further and seek a measure-valued function that can represent the tangent vector. To this author's knowledge, the theory for this has not been fully established, but, we effectively try to encompass the system's sensitivity to perturbation as a Gateaux-like derivative \cite{Gateaux}: \begin{equation} \lim_{\eta \rightarrow 0} \frac{u^{p,s} - u^{0,s}}{\eta} = \begin{cases} \frac{-1}{(1-t)^2}, \quad x < x_s^0(t) \\ 1, \quad x > x_s^0(t) \end{cases} + M(t) \delta(x - x_s^0(t)) \label{gat} \end{equation} with \begin{equation} M(t) \equiv \lim_{\eta \downarrow 0} \frac{1}{\eta} \int_{x_s^p(t)}^{x_s^0(t)} ( \frac{\eta}{1 - \eta t} - \frac{1}{1-t} ) \dd x = \frac{-t^2}{2(1-t)^2} = -[u^{0,s}]_{x_s^0} \hat{x}, \label{burgermass} \end{equation} encompassing the same information as in (\ref{funVec}) and (\ref{shockVec}).

Here, $\delta$ denotes the Dirac-delta function \cite{Evans}. 
This is a rather natural interpretation for the scenario, that a small segment of area on which the base and perturbed solutions have an $O(1)$ difference as $\eta$ tends to zero is captured by concentrating that difference in the form of a delta function localized to a single point. More discussion follow below.

\subsubsection{Linear Perturbation}

We now turn to the question of how a linear perturbation affects a hyperbolic system. In particular, we are interested in describing the time evolution of the difference in shock positions between the base and perturbed solutions. 
 The perturbation of hyperbolic systems is quite delicate. We begin this section with a brief overview of the established theory, from which we will infer equations suitable for our purposes. 

We begin with \begin{equation} {\bf u}_t + {\bf f(u)}_x = {\bf g(u)} \label{baseVec} \end{equation}
subject to the perturbation of initial values ${\bf u}(x,t=0;\eta) = {\bf u_0}(x,t=0;\eta) + \eta {\bf v}(x,t=0;\eta)$ where $\eta \ll 1.$ The naive linearized system with ${\bf u}(x,t) = {\bf u_0}(x,t; \eta) + \eta {\bf v}(x,t; \eta) + O(\eta^2)$ producing \begin{align}
{{\bf u}_{{\bf 0}}}_{,t} + {\bf f}({\bf u_0})_x &= {\bf g(u_0)} \label{linO1} \\
{\bf v}_t + (A({\bf u_0}) {\bf v})_x &= B({\bf u_0}) {\bf v} \label{linOeta}
\end{align} where $A({\bf u_0})$ and $B({\bf u_0})$ are the Jacobian matrices $\frac{\partial {\bf f}}{\partial {\bf u}} ({\bf u_0})$ and $\frac{\partial {\bf g}}{\partial {\bf u}} ({\bf u_0})$, respectively, turns out to be the mathematically valid linearization \cite{linStab}. We will later denote ${\bf u_0^0}$ as the solution when $\eta = 0$, ${\bf u}(x,t;\eta=0)$, and ${\bf v^0}$ as ${\bf v}(x,t;\eta=0)$. In our analysis, we must not mix ${\bf u_0^0}$ up with ${\bf u_0}$, which is the $O(1)$ component of the solution for a given $\eta \neq 0$.

The authors of \cite{linStab} analyze this system, (\ref{linO1}) and (\ref{linOeta}), in the reference frame of the shock, which has theoretical advantages in that in the reference frame of the shock and under a coordinate transformation, the system of the linearized solutions and perturbed shock difference are known to have existence and uniqueness \cite{Majda}. As our numerical work is not done in such a reference frame, we shall not go that route, but the well-posedness of what should amount to the identical system and the strong agreement between the numerics and our theory, serve as support for the systems we develop below to describe the shock front perturbation.

If $\eta$ were 0, we anticipate a shock at position $s_0(t)$ whereby \begin{equation} [{\bf f(u_0^0)}]_{s_0} = s_0'(t) [{\bf u_0^0}]_{s_0}. \label{basicS0} \end{equation} We are explicit here by placing $s_0$ in the subscript of the [] that the discontinuity is at $s_0.$ In general the presence of a perturbation will cause the shock to move at a slightly different velocity and also change the value of the solution. We therefore choose an asymptotic expansion to represent the Rankine-Hugoniot condition and write $s \sim s_0 + \eta s_1 + ...$ giving \begin{align}
[{\bf f}({\bf u_0} + \eta {\bf v} + ...)]_{s_0 + \eta s_1 + ...} &= (s_0'(t) + \eta s_1'(t) + ...)[{\bf u_0} + \eta {\bf v} + ...]_{s_0 + \eta s_1 + ...} \implies \nonumber \\
[{\bf f}({\bf u_0}) + \eta A({\bf u}) {\bf v} + ...]_{s_0 + \eta s_1 + ...} &= (s_0'(t) + \eta s_1'(t) + ...)([{\bf u_0}]_{s_0 + \eta s_1 + ...} + \eta [{\bf v}]_{s_0 + \eta s_1 + ...} + ...) \implies \nonumber \\
[{\bf f}({\bf u_0^0})]_{s_0} + \eta [A({\bf u_0^0}) {\bf u_0^0}_{,x}]_{s_0} s_1 + \eta [A({\bf u_0}) {\bf v}]_{s(t)} + ... &= s_0'(t) [{\bf u_0^0}]_{s_0} \nonumber \\ & + \eta s_0'(t) s_1(t) [{\bf u_0^0}_{,x}]_{s_0} + \eta s_0'(t) [{\bf v}]_{s(t)} + \eta s_1'(t) [{\bf u_0^0}]_{s_0} + ... \nonumber \implies \\
[{\bf f}({\bf u_0^0})]_{s_0} + \eta [A({\bf u_0^0}) {\bf u_0^0}_{,x}]_{s_0} s_1 + \eta [A({\bf u_0^0}) {\bf v^0}]_{s_0} + ... &= s_0'(t) [{\bf u_0^0}]_{s_0} + \eta s_0'(t) s_1(t) [{\bf u_0^0}_{,x}]_{s_0} + \eta s_0'(t) [{\bf v^0}]_{s_0} + \eta s_1'(t) [{\bf u_0^0}]_{s_0} + ... \label{shockLin0}
\end{align}
Because both ${\bf u_0}$ and ${\bf u_0^0}$ represent $O(1)$ contributions to the solution, apart from the small region of $O(\eta)$ brought about by a small change in shock speeds, ${\bf u_0}$ and ${\bf u_0^0}$ should be equal. Knowing that ${\bf u_0^0}$ suffers a discontinuity at $s_0$ but is otherwise continuous (and we assume smooth) elsewhere, we smoothly extrapolate ${\bf u_0^0}(s_0^\pm,t)$ to estimate ${\bf u_0}(s(t)^\pm,t;\eta)$ in our work to derive (\ref{shockLin0}). In much of the theory theory, the solutions considered are taken to be piecewise (away from the shock) Lipschitz continuous \cite{Gateaux}, giving differentiability almost everywhere. Also pertinent to the derivation of (\ref{shockLin0}) is that at leading order, $[A({\bf u_0}) {\bf v}]_{s(t)}$ and $[{\bf v}]_{s(t)}$ are respectively, $[A({\bf u_0^0}) {\bf v^0}]_{s_0}$ and $[{\bf v^0}]_{s_0}$. Equation (\ref{shockLin0}) will be relevant shortly.

It is important to note that in our physical situation, we can compute ${\bf u_0^0}$, i.e., the solution without a perturbation, but we are not computing ${\bf u_0}$, the $O(1)$ component to the solution with the perturbed shock position. In turn, this will only afford us a linearized system ${\bf u}(x,t) = {\bf u_0^0}(x,t) + \eta {\bf w}(x,t) + ...$ and given the form of (\ref{linO1}) and (\ref{linOeta}), we posit that
\begin{align}
{{\bf u_0^0}}_{,t} + {\bf f}({\bf u_0^0})_x &= {\bf g(u_0^0)} \label{linO1num} \\
{\bf w}_t + (A({\bf u_0^0}) {\bf w})_x &= B({\bf u_0^0}) {\bf w}, \label{linOetanum}
\end{align}
valid away from the shock at $s_0(t)$ obeying (\ref{basicS0}).

Given that ${\bf u}(x,t;\eta) = {\bf u_0}(x,t;\eta) + \eta {\bf v}(x,t;\eta) + ...$, we compute 
\begin{align} {\bf w}(x,t) &= \lim_{\eta \rightarrow 0} \frac{{\bf u}(x,t;\eta) - {\bf u_0^0}(x,t)}{\eta} \nonumber \\
&= \lim_{\eta \rightarrow 0} (\frac{{\bf u_0}(x,t;\eta) - {\bf u_0^0}(x,t)}{\eta} + {\bf v}(x,t;\eta) + ...) \nonumber \\
&= -[{\bf u_0^0}]_{s_0(t)} s_1(t) \delta(x(t) - s_0(t)) + {\bf v}^0 \nonumber
\end{align}
to be the linearized perturbation. The first term of the line above should be interpreted in a weak sense as $\eta \rightarrow 0$, and the second term is a pointwise (and also weak) limit of ${\bf v}(x,t;\eta)$ as $\eta \rightarrow 0$. As $\eta \rightarrow 0$, the discontinuity of ${\bf v}$ approaches $s_0(t)$, with ${\bf v}$ being a smooth function away from the discontinuity.

From a numerical perspective, we therefore anticipate that ${\bf w}$ will be a combination of a function that is smooth on either side of $s_0$ and a measure-valued function located at $x=s_0(t)$. Outside of the small numerical region holding the delta function, ${\bf w} = {\bf v}$. We now turn our attention to deriving an ODE for the size of the delta function.

From (\ref{shockLin0}), from the $O(1)$ and $O(\eta)$ components of the system, we have 
\begin{align} [ {\bf f}({\bf u_0^0}) ]_{s_0} &= s_0'(t) [{\bf u_0^0}]_{s_0} \label{ord0Lin} \\ 
[(A({\bf u_0^0}) - s_0'(t)) {\bf v^0}]_{s_0} &= s_1'(t) [{\bf u_0^0}]_{s_0} + s_1(t) [ (s_0'(t) - A({\bf u_0^0})) {\bf u_0^0}_{,x}]_{s_0}. \label{ord1Lin}
\end{align}
Denoting ${\bf M(t)} = - s_1(t) [{\bf u_0^0}]_{s_0}$, then $$s_1'(t) [{\bf u_0^0}]_{s_0} = -{\bf M}'(t) - s_1(t) \frac{\dd}{\dd t} [{\bf u_0^0}]_{s_0(t)} = -{\bf M}'(t) - s_1(t) ( [{\bf u_0^0}_{,t}]_{s_0} + s_0'(t) [{\bf u_0^0}_{,x}]_{s_0} )$$ and ${\bf u_0^0}_{,t} = {\bf g}({\bf u_0^0}) - A({\bf u_0^0}) {\bf u_0^0}_{,x},$ away from $x=s_0(t)$, so we conclude \begin{equation} {\bf M}'(t) = -[(A({\bf u_0^0}) - s_0'(t) ) {\bf v^0}]_{s_0} + s_1(t) [{\bf g}({\bf u_0^0})]_{s_0} \label{matMass} \end{equation} and that in component form, where subscript $i$ denotes the $i$th vector component, when $([{\bf u_0^0}]_{s_0})_i \neq 0$, \begin{equation} M_i'(t) = -([(A({\bf u_0^0}) - s_0'(t)) {\bf v^0}]_{s_0})_i + \frac{[{\bf g}({\bf u_0^0})]_i}{([{\bf u_0^0}]_{s_0})_i} M_i(t). \label{Mvec} \end{equation} 

Having an equation of the form (\ref{Mvec}) is useful because in terms of well-known properties of the unperturbed solution ${\bf u_0^0}$, we have an ordinary differential equation for the size of the region over which the base and perturbed solutions differ in their $O(1)$ behaviour, i.e., the mass of the delta function.

\subsection{Linear Perturbation Example}

We can now question how a linear perturbation to (\ref{GM}) as per (\ref{ICp}) behaves. We can consider this as a problem with $u(x,t) = u^{0,s}(x,t) + \eta u^{1,s} (x,t)$ with \begin{equation} u^{1,s}(x,0) = \begin{cases} -1, \quad x < 0 \\ 1, \quad x \geq 0 \end{cases} \label{linic} \end{equation} and with $u^{1,s}$ obeying the linearized (about $u^{0,s}$) problem: \begin{equation} u^{1,s}_{t} + (u^{0,s} u^{1,s})_x = 2 u^{0,s} u^{1,s} \label{GMlin} \end{equation}

Here the characteristic speeds for $u^{1,s}$ are $u^{0,s}$, i.e., $\frac{1}{1-t}$ and $0$, and along the characteristic curves, either $$\frac{\dd u^{1,s}}{\dd t} = \frac{2}{1-t} u^{1,s}, u^{1,s}(0) = -1 \implies u^{1,s} = \frac{-1}{(1-t)^2} \quad \text{left of the shock}$$ or $$\frac{\dd u^{1,s}}{\dd t} = 0, u^{1,s}(0) = 1 \implies u^{1,s} = 1 \quad \text{right of the shock}.$$ Noting this regular part of the solution is not difficult, but we must have that $$u^{1,s}(x,t) = \begin{cases} \frac{-1}{(1-t)^2}, \quad x < x_s^0(t) \\ 1 \quad x > x_s^0(t) \end{cases} + M(t) \delta(x-x_s^0(t)).$$ Indeed, there must be a moving delta function for any solution, even weak, to exist. The flux function $u^{0,s} u_1$ is $0$ on the right with $u^{0,s} = 0$, but there is a nonzero flux on the left as $u^{0,s} u_1 = \frac{-1}{(1-t)^3}.$ This imbalance in flux arises from the fact the shock moves at a speed prescribed by $u^{0,s}$, not by a flux balance for $u^{1,s}.$

Using (\ref{Mvec}), applying it to this scalar case, we have \begin{align*} \frac{\dd M}{\dd t} &= -[(u^{0,s} - \frac{1}{2(1-t)}) u_1]_{s_0} + \frac{[(u^{0,s})^2]_{s_0}}{[u^{0,s}]_{s_0}} M \\ &= \frac{1}{2(1-t)} - \frac{1}{2(1-t)^3} + \frac{1}{1-t} M. \end{align*} With $M(0) = 0$, this can be solved with an integrating factor to yield $M(t) = \frac{-t^2}{2(1-t)^2}$, in agreement with the delta-function in the Gateau-derivative, (\ref{gat}) and (\ref{burgermass}). 


\subsection{Perturbed Burger Equation - Numerical Results}

By using a first-order finite volume upwind scheme on a uniform mesh on $-1 \leq x \leq 1$ with spatial mesh size $h \propto 1/N$, and with a split-step in time to manage geometric sources \cite{Randy}, we look at the results of solving 
(\ref{GMlin}) with (\ref{linic}). The upwinding in this situation is trivial as we know the characteristics are always moving to the right or stationary for the nonlinear system and the perturbed system alike. We remark, however, that care is needed in how the systems are solved. We solve the $u_0$ advancement separately from the $u_1$, and solve both with a split-step: first, we advance $u_0$ based on its flux; then, we advance $u_1$ based on its flux with the $u_0$-value prior to its update; then, we update $u_0$ with its source term; finally, we update $u_1$ with its source term using the value of $u_0$ prior to its source term advancement. Using the original $u_0$-value for all of the $u_1$-steps fails to obtain the correct mass of the delta function.

\begin{figure}
 \begin{center}
    \includegraphics[width=5.in]{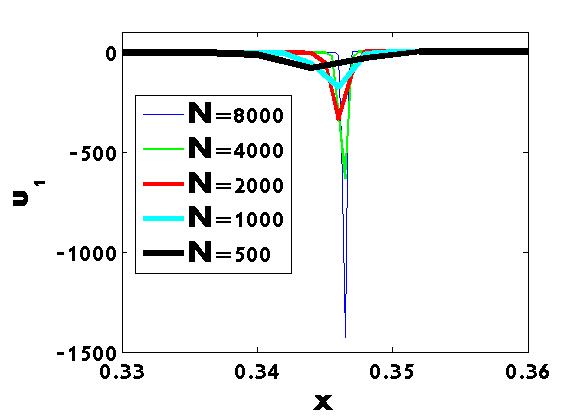}
 \end{center}
\caption{The figure depicts the numerically computed solution $u_1$ as a function of $x$ at time $t=1/2$ for $N=500,1000, 2000, 4000$ and 8000. We can observe the numerical emergence of a delta function.}
\label{f:BurgerDeltas}
\end{figure}

Figure \ref{f:BurgerDeltas} depicts the numerical solution $u_1$ using this simple first-order scheme for different values of $N$. As $N$ increases, the spiked region, the delta function, becomes narrower and narrower and the height grows. Table \ref{tab:massconv} documents the numerically integrated masses $M_{ODE}$ of the delta function using an ODE advancement based on equation (\ref{Mvec}) and based on a first-order numerical integration of the support of the delta-function $M_{Integral}$. The ODE jump discontinuities are measured by taking values far to the left and far to the right of the shock. The support is established by finding the x-value where $u_1$ becomes positive and integrating $u_1(x,t) - u_1(-1,t)$ up to this x-value, the idea being that the solution is constant up to a small transition region and once $u_1$ is positive, the delta function has been passed. 

The exact value of the delta function coefficient at $t=0.5$, the stopping time of our simulations, should be $-0.5 = M(0.5)$. Both methods yield convergent results. As the numerical scheme is first-order, we can also extrapolate the values predicted: if $h$ is the mesh size then we anticipate that $M(h) = M^* + Ah + o(h)$ where $M(h)$ denotes the delta function mass prediction at size $h$, $M^*$ is the exact value, $A$ is a constant that gives the $O(h)$-component of the error, and $o(h)$ is a yet smaller correction. By extrapolating the schemes' results to $h=0$ with a linear fit, we also tabulated the extrapolated values, which are far more accurate. The data are found in table \ref{tab:massconv}. We remark that this argument isn't entirely obvious and these ideas constitute an educated guess and not a definitive conclusion on the numerical convergence to the delta-function strength: hyperbolic numerical schemes converge in the $L^1$-norm, not pointwise. Choosing a small window near the delta-function and having first-order convergence in $L^1$ should intuitively mean the mass of the delta function has error that is $O(h)$. If the value of the numerical solution on either side of the jump discontinuity were not within $O(h)$ of its value then we couldn't hope to have first-order convergence in $L^1$ in a region near but not straddling the shock. We also wish to emphasize that these results are rather clean but in general, based on how the shocks are smeared out due to the numerical scheme, it can be very difficult to measure or predict how convergence will take place \cite{Randy}.

\begin{table}[h]
\begin{center}
\begin{tabular} {c | c | c | c | c | c | c}
$N$ & $M_{ODE}$ & Error & $M_{Integral}$ & Error \\
\hline
500 & -0.4976 & 0.0024 & -0.4443 & 0.0557 \\
1000 & -0.4988 & 0.0012 & -0.4685 & 0.0315 \\
2000 & -0.4994 & 0.0006 & -0.4856 & 0.0144 \\
4000 & -0.4997 & 0.0003 & -0.4918 & 0.0082 \\
8000 & -0.4998 & 0.0002 & -0.4961 & 0.0039 \\
\hline
Extrapolated & -0.5 & 0.0000 & -0.4988 & 0.0012
\end{tabular} 
\end{center}
\caption{Convergence of the numerically estimated delta-function mass to the analytic value of $-0.5$.} \label{tab:massconv}
\end{table}

We similarly validate the consistency of the numerically computed delta function mass and ODE-predicted delta function mass under a more complex problem situation $$u_t + (\frac{1}{2} u^2)_x = \sqrt{u},$$ with $$u(x,0) = \begin{cases} -2x+10, \quad x<0 \\ 1 - x/20, \quad x > 0 \end{cases} + \eta \begin{cases} -1, \quad x < 0 \\ 1, \quad x > 0. \end{cases}$$ 
The numerically computed mass of the delta function at $t=0.08$ is $0.0292$ and with the ODE (\ref{Mvec}), we predict $0.0293$. 


\subsection{Interpretation}

From our analysis and prior test cases, we have that the mass of the delta function is the negative jump in the base solution, $[{\bf u_0^0}]_{s_0}$ times $s_1(t)$, the first-order correction to the shock position (normalized by $\eta$). If $\Delta_s(\eta, t)$ denotes the difference in front positions (perturbed minus base) then 

\begin{equation} {\bf M}(t) = - [{\bf u_0^0}] \partial_\eta \Delta_s(\eta,t) | _{\eta=0}. \label{Delprime} \end{equation} In general, we can infer the spatial distance over which the base and perturbed solutions significantly differ based on knowing the jump in the base solution and the mass of the delta function.

\section{Asymptotic Study of High Azimuthal Numbers}

In this section we study a toy model of an implosion that provides understanding into the implosion process for various extreme scenarios, and we ask the question: can high azimuthal mode numbers cause instabilities during the formation of pressure pulses? Through this analysis, we also learn about how the numerics may (and do) behave in solving the nonlinear system at high mode numbers and foreshadow some of the challenges in the simulations.

\label{Asy}

\subsection{Asymptotic Insights from within the Linear Acoustic Limit}


From the asymptotic analysis of a similar model with spherical symmetry \cite{asy}, the leading order equations amounted to those describing linear acoustics, namely:

\begin{equation} \rho_t + \nabla \cdot \vv = 0 \label{linacmass} \end{equation} \begin{equation}\vv_t + c^2 \nabla \rho = \zz \label{linacmom} \end{equation} in a suitably scaled coordinate regime. Here $\rho$ denotes a perturbation from a baseline density, $\vv$ is the leading order velocity, $t$ is a fast time scale and $c$ is a rescaled sound speed. We consider these equations holding within a dimensionless annulus $a < r < 1$.

Taking the divergence of (\ref{linacmom}) we find that $c^2 \bigtriangleup \rho = -(\nabla \cdot \vv)_t$ and from taking a time derivative of (\ref{linacmass}) and using the result we have that \begin{equation} \rho_{tt} = c^2 \bigtriangleup \rho. \label{rhowave} \end{equation} In this regime, the density satisfies the wave equation. As the equation is linear, we pose that $\rho = u(r,t) \e^{im\theta}$ to describe a linear perturbation subject to (\ref{rhowave}) with \begin{align} u(r,0) = u_t (r,0) &= 0 \label{linacic} \\ u(1,t) = f(t) &= \e^{-t^2} \label{linacoutbc} \\ u(a,t) &= 0. \label{linacinbc} \end{align}

One question that yields considerable insight is how perturbation behaves if $m$ is very large: this could describe the growth of instabilities for high azimuthal mode numbers, which are likely the only ones present in a well-engineered reactor. Equations (\ref{rhowave}) to (\ref{linacinbc}) can be solved exactly with Fourier series, however there are two limits involved: one describing the mode number and the other, the number of terms in the series sum being used. In order to gain insight into the asymptotic nature of the pulse formation for $m \gg 1$ in a uniformly valid regime where we do not need to worry when the term number in the series is below or above $m$, we will solve the problem asymptotically. Due to the combination of multiple asymptotic limits, namely the high mode numbers and those that reduce the system to the linear acoustic equations, we believe the solutions obtained through this analysis are insightful, but may only be weakly qualitative in the physical system.
As the mode numbers are our primary concern with other parameters $O(1)$, for simplicity rescale $t$ to $t \rightarrow t/c$ and consider
\begin{equation}
u_{tt} = u_{rr} + \frac{1}{r} u - \frac{1}{m^2} u, \quad u(r,0) = u_t(r,0) = 0, u(1,t) = f(ct)
\end{equation}
where $f(t)$ represents the impulse. As $m \rightarrow \infty,$ with $m (1-r) = x,$ $T = mt$ and with $u \sim u_0 + m^{-1} u_1 + ...$, at leading order (with $t$ replacing $T$), we obtain:

\begin{align}
u_{0,tt} &= u_{0,xx} - u_0 \quad (x,t) \in [0, \infty) \times [0, \infty) \label{asyeq}  \\
u_0(x,0) &= u_0(0,t) = 0, u(0,t) = f(0) = 1 \label{asybcs} 
\end{align}

This amounts to the Linear Klein-Gordon equation in one space dimension. One strategy that has been fruitful in solving this equation for other geometries and boundary conditions is the technique of Adomian Decomposition \cite{ado}, however the method does not work here due to the discontinuity emerging from $(x=0,t=0).$ To solve equations (\ref{asyeq}) and (\ref{asybcs}), we solve a similar problem on $(x,t) \in [0, R] \times [0, \infty)$, $R>0$, with: \begin{align*} \tilde{u}_{tt} &= \tilde{u}_{xx} - \tilde{u} \\ \tilde{u}(x,0) &= \tilde{u}_t(x,0) = 0 \\ \tilde{u}(0,t) &= 1, \tilde{u}(R,t) = 0 \end{align*}

By writing $\tilde{u}$ as a sum of a particular solution $\cosh x - \coth R \sinh x$ that is time-independent and an equation with homogeneous spatial boundary conditions $\tilde{v}$ with
\begin{align*} \tilde{v}_{tt} &= \tilde{v}_{xx} - \tilde{v} \\ \tilde{v}(x,0) &= - \cosh x + \coth R \sinh x \\ \tilde{v}_t(x,0) &= 0 \\ \tilde{v}(0,t) &= \tilde{v}(R,t) = 0, \end{align*}
we can write $$\tilde{u}(x,t) = \cosh x - \coth R \sinh x + \sum_{n=1}^{\infty} \frac{-2}{R} \left( \int_0^R h_R(y) \sin(\lambda_{n,R} y) \dd y \right) \sin(\lambda_{n,R} x) \cos(\sqrt{1 + \lambda_{n,R}^2} t)$$ where $\lambda_{n,R} = \frac{n\pi}{R}$, and $h_R(y) = \cosh y - \coth R \sinh y.$ In the limit as $R \rightarrow \infty$, following the derivation of the Fourier Transform in \cite{saffsnider}, we obtain that $\tilde{u} \rightarrow u_0$ with \begin{equation} u_0(x,t) = \exp(-x) - \frac{2}{\pi} \int_0^{\infty} \frac{\omega}{1 + \omega^2} \sin(\omega x) \cos(\sqrt{1 + \omega^2} t) \dd \omega. \label{usolint} \end{equation}

The integral in (\ref{usolint}) closely resembles many of the Bateman integrals \cite{bateman} arising in Fourier sine and cosine transforms; however, this particular integral is not documented and it seems it cannot be evaluated analytically except for special cases. When $x \geq t$, we shall observe that $u(x,t)$ can be evaluated exactly. When $x < t$, we will obtain a leading-order asymptotic form for the integral for large $t$.

Let \begin{align} I &= \int_0^{\infty} \frac{\omega}{1 + \omega^2} \sin(\omega x) \cos(\sqrt{1 + \omega^2} t) \dd \omega \nonumber \\ &= \frac{1}{2} \text{Im} \int_{-\infty}^{\infty} \frac{\omega}{1 + \omega^2} \e^{i \omega x} \cos(\sqrt{1 + \omega^2} t) \dd \omega \nonumber \\ &=   \frac{1}{4} \text{Im} \int_{-\infty}^{\infty} \left( \underbrace{ \frac{\omega}{1 + \omega^2} \e^{i (\omega x + \sqrt{1 + \omega^2} t) } }_{J^+} + \underbrace{ \frac{\omega}{1 + \omega^2} \e^{i ( \omega x-\sqrt{1 + \omega^2} t) } }_{J^-} \right) \dd \omega \label{theI} 
\end{align}

\begin{figure}
 \begin{center}
    \includegraphics[width=2.5in]{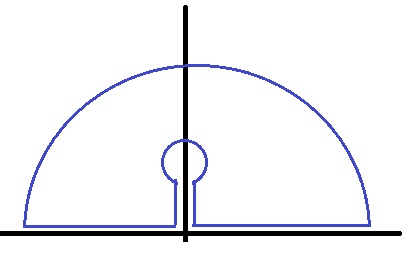}
 \end{center}
\caption{Sketch of the chosen contour.}
\label{f:contours}
\end{figure}

To evaluate $\int_{-\infty}^{\infty}$, we use contour integration and write $$(\int_{\gamma^-} + \int_{\gamma^\ell} + \int_{\gamma^\epsilon} + \int_{\gamma^r} + \int_{\gamma^+} + \int_{\gamma^R}) J^\pm \dd \omega = 0$$ where 
\begin{align*}
\gamma^-&: \quad \omega = s + 0^+ i, -\infty < s < 0 \\
\gamma^\ell&: \quad \omega = 0^- + is, 0 < s < 1 - \epsilon \\
\gamma^\epsilon&: \quad \omega = i + \epsilon \e^{i \theta}, 3 \pi/2 > \theta > - \pi/2 \\
\gamma^r&: \quad \omega = 0^+ + is, 1 - \epsilon > s > 0 \\
 \gamma^+&: \quad \omega = s + 0^+ i, 0 < s < \infty \\
 \gamma^R&: \quad \omega = R \e^{i \theta}, 0 < \theta < \pi \\
\end{align*} in such a way as the contours line up (we must take $\epsilon \downarrow 0$ and $R \rightarrow \infty$). See figure \ref{f:contours}. We define our branch of the square root function here with $$\sqrt{1 + \omega^2} = |1 + \omega^2|^{1/2} \e^{\frac{1}{2} ( \text{Arg}_{-\pi/2} (1 + i \omega) + \text{Arg}_{-\pi/2} (1 - i \omega) ) }$$ where $\text{Arg}_{-\pi/2} \in [-\pi/2, 3\pi/2).$ Such a choice of branch induces a branch cut along $is$ with $-1 \leq s \leq 1.$ This will be necessary for the contour integration. We begin by assuming $x > t$.

In this case, \begin{align} ( \int_{\gamma^-} + \int_{\gamma^+}) J^\pm \dd \omega &= \int_{-\infty}^{\infty} J^\pm \dd \omega \\
\int_{\gamma^\epsilon} (J^+ + J^-) \dd \omega &= \int_{3\pi/2}^{-\pi/2} \frac{i + O(\epsilon)}{2i \epsilon \e^{i \theta} + O(\epsilon^2)} ( \e^{i(ix + O(\sqrt{\epsilon}))} + \e^{i(ix - O(\sqrt{\epsilon}))} ) i \epsilon e^{i \theta} \dd \theta \nonumber \\
& \rightarrow -2 \pi i \e^{-x}, \quad \epsilon \downarrow 0 \label{sqtpt} \\ \int_{\gamma^R} (J^+ + J^-) \dd \omega &= \int_0^{\pi} \frac{R \e^{i \theta}}{1 + R^2 \e^{2 i \theta}} (\e^{i(R\e^{i \theta} x + R\e^{i\theta} t(1 + O(1/R^2)))} \nonumber \\ &+ \e^{i(R\e^{i\theta}x - R\e^{i\theta}t(1 + O(1/R^2)))}) i R \e^{i \theta} \dd \theta \nonumber \\ &\sim \int_0^\pi i ( \e^{-(x+t)R\sin \theta} + \e^{-(x-t)R\sin \theta} ) \times O(1)  \dd \omega \nonumber \\ & \rightarrow 0, \quad R \rightarrow \infty  \label{lgpt} \\ ( \int_{\gamma^\ell} + \int_{\gamma^r}) (J^+ + J^-) \dd \omega &= \int_0^1 \frac{is}{1 - s^2} ( (\e^{i(isx - \sqrt{1-s^2}t)} + \e^{i(isx + \sqrt{1-s^2}t)} ) \nonumber \\ &- ( \e^{i(isx + \sqrt{1-s^2}t)} + \e^{i(isx - \sqrt{1-s^2}t)} ) ) i \dd s \nonumber \\ &= 0 \label{cancels} \end{align}

\noindent where in arriving at (\ref{sqtpt}), we use $\sqrt{1 + \omega^2} = O(\sqrt{\epsilon})$ for $\omega = i + O(\epsilon)$; in arriving at (\ref{lgpt}), we note that if $x>t$ then both $x+t$ and $x-t$ are positive so the integrand vanished as $R \rightarrow \infty$ where the $O(1)$ component represents the phase of modulus $1$; and in arriving at (\ref{cancels}), we make use of the branches of the square root function having opposite signs on opposite sides of the imaginary axis and combined the integrals $s = 0$ to $s=1$ and $s=1$ to $s=0$ into a single integrand with $s$ ranging from $0$ to $1$.

From equations (\ref{theI}) through (\ref{cancels}), we find that for $x>t$, that $I = 2 \pi \e^{-x}$ and thus by (\ref{usolint}), \begin{equation*} u = \e^{-x} - \frac{2}{\pi} \frac{1}{4} (2 \pi \e^{-x}) = 0.  \end{equation*} This makes intuitive sense as for $x > t,$ no disturbances could have reached the point $(x,t)$ when limited to the scaled sound speed of $1.$ If $x=t$ then all of the analysis still applies, but because $x-t=0$ we find that (\ref{lgpt}) amounts to 
\begin{equation} \int_{\gamma^R} (J^+ + J^-) \dd \omega = \int_0^\pi i e^0 = \pi i, \quad (R \rightarrow \infty) \label{xet} \end{equation}
giving $I = 2 \pi i \e^{-x} - \pi i$ so if $x=t$ then $u_0 = 1/2$.

For $x<t$, the situation is difficult as the integrals as $R \rightarrow \infty$ only converge with $x+t$ but diverge with $x<t$ because $x-t < 0$. While considering two contours, one which goes around $+i$ and the other which goes around $-i$ may seem appropriate, the integrands that would be necessary to arrive at an equation such as (\ref{cancels}) in fact do not cancel. Instead, one obtains that \begin{align} ( \int_{\gamma^-} + \int_{\gamma^+}) (J^+ + J^-) \dd \omega &= \int_0^1 \frac{is}{1 - s^2} ( \e^{i(isx - \sqrt{1 - s^2} t)} - \e^{i(isx + \sqrt{1 - s^2} t)} \nonumber \\ &- \e^{i(-isx - \sqrt{1 - s^2} t)} + \e^{i(-isx + \sqrt{1 - s^2} t)}  ) i \dd s \nonumber \\ &= 2i \int_0^1 (\frac{s \sin(\sqrt{1-s^2}t)}{1 - s^2}(e^{sx}  - e^{-sx} ) \dd s \nonumber \\ &= 2i \int_0^1 (\frac{\sin(wt)}{w} (\e^{\sqrt{1-w^2} x} - \e^{-\sqrt{1+w^2} x}) \dd w \nonumber \end{align} This integral does not yield to obvious analytic techniques. By going back to (\ref{usolint}), we will seek a leading-order asymptotic expansion for $t \rightarrow \infty$, representing the net growth of the perturbations far away from the outer boundary in this asymptotic regime. A decay in size would indicate stability; growth could indicate instability -- within the linearized regimes.

We denote $$0 \leq \sigma = x/t \leq 1$$ and $$\phi^\pm = t( \sigma \omega \pm \sqrt{1 + \omega^2} )$$ such that $$K^\pm = \int_{-\infty}^{\infty} J^\pm \dd \omega = \int_{-\infty}^{\infty} \frac{\omega}{1 + \omega^2} \e^{i \phi^\pm} \dd \omega.$$ For $0 \leq \sigma < 1 - O(1/t^2)$, the method of stationary phase will be sufficient in obtaining the leading order term. The critical points of $\phi^\pm$ occur where \begin{equation} {\phi^\pm}'(\omega) = t(\sigma \pm \frac{\omega}{\sqrt{1 + \omega^2}}) \label{phip} \end{equation} is zero, which are \begin{equation} \omega^{*\pm} = -\frac{\sigma}{\sqrt{1 - \sigma^2}} \label{omegstar} \end{equation} at which $$\phi^\pm (\omega^{*\pm}) = \pm t \sqrt{1 - \sigma^2},$$ $$(\phi^\pm)'' (\omega^{*\pm}) = \pm t (1 - \sigma^2)^{3/2},$$ and $$\frac{\omega^{*\pm}}{1 + (\omega^{*\pm})^2} = \mp \sigma \sqrt{1 - \sigma^2}.$$ Rescaling according to $\lambda = t (\omega - \omega^{* \pm})$, stationary phase yields that $$K^\pm \sim \mp \sigma \sqrt{1 - \sigma^2} \exp(\pm i t \sqrt{1 - \sigma^2}) \int_{-\infty}^{\infty} \exp(i (1 - \sigma^2)^{3/2} \lambda^2/2) \dd \lambda$$ where in combining $K^+ + K^-$ and evaluating the complex Gaussian integrals, we find 
\begin{equation} u_0(x,t) \sim e^{-x} + \frac{1}{\sqrt{t}} \sqrt{\frac{2}{\pi}} \frac{\sigma}{\sqrt[4]{1 - \sigma^2}} \sin(t \sqrt{1 - \sigma^2} + \pi/4), \quad (x < t). \label{usig0} \end{equation}

This solution is not uniformly valid: as $\sigma = x/t \uparrow 1$, the term that is $O(t^{-1/2})$ changes orders. At $\sigma = 1 - O(1/t^2),$ the term becomes $O(1).$ To deal with this region, we set $\sigma = 1 - \shat/t^2$ with $\shat = O(1)$. In this case, $\omega^{*\pm} = O(t)$ by equation (\ref{omegstar}). Letting $\omega = t \Omega$, with $\Omega = O(1)$ one has that $\int_{-\infty}^{\infty} J^\pm \dd \omega$ from (\ref{theI}) becomes: \begin{align*} \int_{-\infty}^{\infty} J^\pm \dd \omega &= \int_{-\infty}^{\infty} \frac{t \Omega}{1 + t^2 \Omega^2} \e^{ it ((1 - \hat{\sigma}/t^2) t \Omega \pm \sqrt{1 + t^2 \Omega^2})} t \dd \Omega \\ &\sim \int_{-\infty}^{\infty} \frac{1}{\Omega} \e^{it (t \Omega - \frac{\hat{\sigma} \Omega}{t} \pm t |\Omega| (1 + \frac{1}{2 t^2 \Omega^2}) )} \dd \Omega \\ &\sim \begin{cases} \int_{-\infty}^0 \frac{1}{\Omega} \e^{i (-\hat{\sigma} \Omega - \frac{1}{2 \Omega} )} \dd \Omega, \quad \text{ for } J^+ \\ \int_0^\infty \frac{1}{\Omega} \e^{i (-\hat{\sigma} \Omega - \frac{1}{2 \Omega} )} \dd \Omega, \quad \text{ for } J^-. \end{cases} \end{align*}

Above, the two possible cases depending on whether we evaluate $J^+$ or $J^-$ arise from the fact that $J^+$ and $J^-$ have stationary phases for $\Omega < 0$ and $>0$, respectively, and due to the oscillatory nature of the integrals with $O(1)$ oscillations near the stationary phase and $\geq O(t)$ elsewhere, the half-lines where the stationary phases do not occur will be negligible.

Then, $$\int_{-\infty}^{\infty} (J^+ + J^-) \dd \omega \sim \int_{-\infty}^{\infty} \frac{1}{\Omega} \e^{i (-\hat{\sigma} \Omega - \frac{1}{2 \Omega} )} \dd \Omega = \int_0^{\infty} \frac{2}{\Omega} \e^{i (-\shat \Omega - \frac{1}{2 \Omega})} \dd \Omega.$$ In writing $\int_0^\infty = \int_0^1 + \int_1^\infty$ and with the change of variables $\hat{\Omega} = 1/\Omega$, this can be rewritten as $\int_{-\infty}^{\infty} (J^+ + J^-) \dd \omega \sim 2i F(\shat)$ with \begin{equation} F(\shat) := \int_0^1 \frac{-1}{\Omega} ( \sin(\shat \Omega + \frac{1}{2 \Omega}) + \sin(\frac{\shat}{\Omega} + \frac{\Omega}{2}) ) \dd \Omega. \end{equation}

At $\shat = 0,$ the integral $F(\shat)$ can be evaluated exactly by a contour integral yielding $-\pi/2$. This is consistent with the evaluation of (\ref{usolint}) with $x=t$ as found in (\ref{xet}). Otherwise, it is possible to numerically evaluate $F(\shat)$. We remark briefly that $F(\shat)$ does converge for all $\shat$ with a simple proof. 

{\it Theorem:} $\int_0^1 \frac{1}{z} \sin(az + b/z) \dd z$ is convergent for any real numbers $a$ and $b$.

{\it Proof:} 
We transform the integral with $z \mapsto 1/z$, yielding $\int_1^\infty \frac{1}{z} \sin(az + b/z) \dd z$. \\ We begin by assuming $b \neq 0$. For large $z,$ the zeros of the sine function being integrated follow the asymptotics $z_n = n\pi/b + O(1/n)$ such that if the integrand is positive on $(z_n, z_{n+1})$ and negative on $(z_{n+1}, z_{n+2})$ then the net area on $[z_n, z_{n+2}]$ is bounded above by $A_n = (z_{n+2} - z_n) (\frac{1}{z_n} - \frac{1}{z_{n+2}}) = O(1/n^2)$ and since $\sum \frac{1}{n^2}$ converges, so must the integral. \\
On the other hand if $b=0$ then $\int_0^1 \frac{1}{z} \sin(az) \dd z$ converges as the integrand is defined and bounded except at the isolated point $z=0$ and the range of integration is finite $\square$.

\begin{figure}
 \begin{center}
    \includegraphics[width=3.25in]{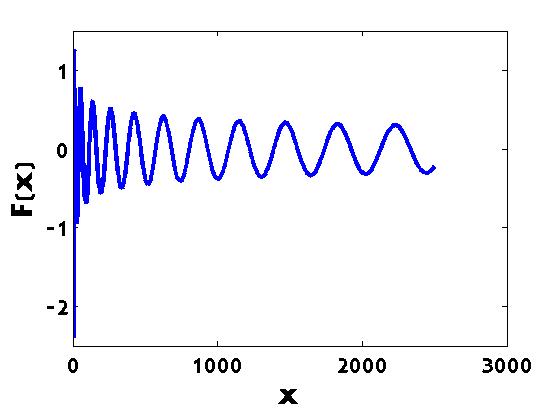}
 \end{center}
\caption{Numerically integrate function $F(x)$.}
\label{f:Fplot}
\end{figure}

We implement a very simple numerical methodology as the integral is a secondary result to the main focus of this section, namely the asymptotic limit and its physical significance. We map $w \in [0,1]$ to $\shat \in [\epsilon, 1]$ with $\shat(0) = \epsilon$, $\shat(1) = 1$, such that $\frac{\dd \shat}{\dd w} \propto \frac{1}{\frac{a}{\shat^2} + b}$, where the denominator is the asymptotic scaling of the derivative of the argument of sine as $\shat \downarrow 0$. This places more mesh points in the region where the integrand is largest and changing most rapidly. The result turns out to be independent of $a,$ and $b$ giving that $\shat(w) = \frac{1}{(1 - 1/\epsilon)w + 1/\epsilon}$. By integrating with the trapezoidal rule in the non-uniform $\shat$-meshing with a C++ program, we calculate our results and plot the function $F(\shat).$ A plot of $F$ is given in figure \ref{f:Fplot}.

We can write this solution valid for $x/t \uparrow 1$ for large $t$ as \begin{equation} u_0 \sim \e^{-x} - \frac{1}{\pi} F(\shat). \label{uFrelate} \end{equation}

\begin{figure}
 \begin{center}
  $\begin{array}{cc}
    \includegraphics[width=3.25in]{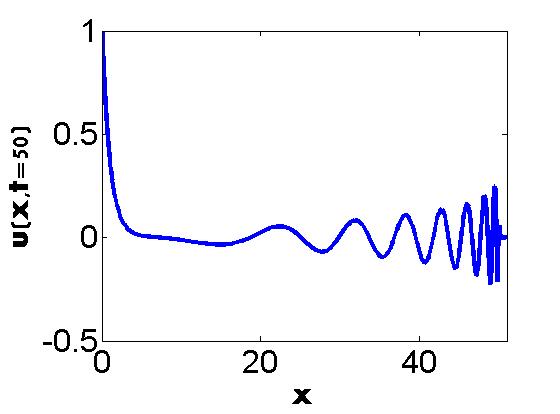} &
\includegraphics[width=3.25in]{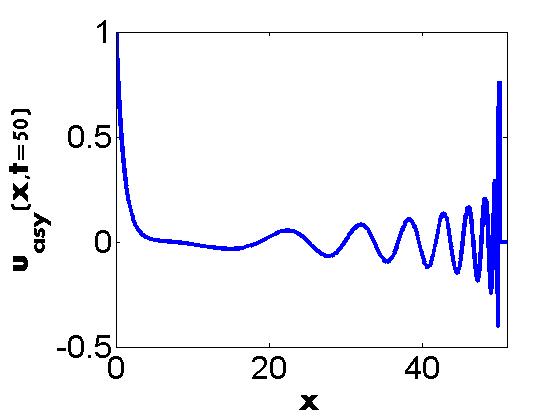}
   \end{array}$
 \end{center}
\caption{Numerical solution at $t=50$ on the left and the asymptotic solution at $t=50$ on the right. The plots are nearly identical except for near $x=50$ where there appears to be a less-defined peak in the numerics, likely due to numerical dissipation.}
\label{f:numasybigt}
\end{figure}

\begin{figure}
 \begin{center}
    \includegraphics[width=3.25in]{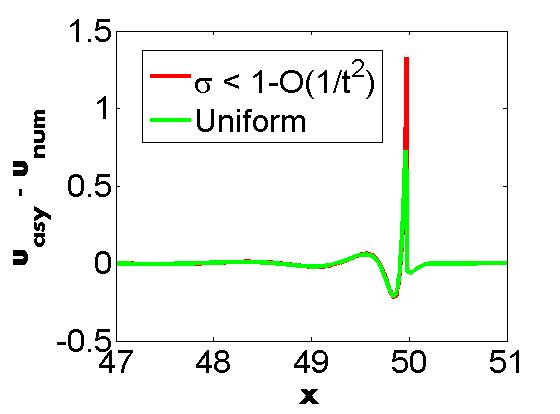}
 \end{center}
\caption{The errors attributed to the asymptotic solution that extends $x/t < 1-O(1/t^2)$ to the entire interval $[0,1]$ and the uniformly vaid asymptotic solution. There is clearly numerical dissipation present as the analytic solution is $0$ for $t>50$ and the asymptotic result is exact here, but the value of the difference is nonzero for $x>t=50$.}
\label{f:asyimps}
\end{figure}

To furnish a uniformly valid asymptotic solution, we shall refer to the solution just obtained as the outer solution and the solution obtained in (\ref{usig0}) as the inner solution. We express both equations in the outer solution coordinates, involving $\shat$: adding the solutions together and subtracting the overlap region defined by $\sigma \uparrow 1$ in the inner solution, obtaining:

\begin{equation} u_0 \sim \begin{cases} \e^{-x} + \sqrt{\frac{2}{\pi}} \left( \frac{1 - \shat/t^2}{\sqrt[4]{2 \shat + \shat^2/t^2}} \sin( \sqrt{2 \shat + \shat^2/t^2} + \pi/4) - \frac{1}{\sqrt[4]{2 \shat}} \sin(\sqrt{2 \shat} + \pi/4) \right) - \frac{1}{\pi} F(\shat), \quad x<t \\ \frac{1}{2}, \quad x=t \\ 0, \quad x > t \end{cases} \label{uuniform} \end{equation} with exactness for $x \geq t.$ The uniformly valid asymptotic approximation is plotted next to the numerical solution at $t=50$ in figure \ref{f:numasybigt}. Due to the fact that the numerics are unable to perfectly pinpoint the location of the shock, it is difficult to validate the asymptotics in the region $x/t = 1 - O(1/t^2)$. We do find evidence, however, in plotting how the $x/t < 1-O(1/t^2)$ asymptotic solution (that has a vertical asymptote) and the uniformly valid asymptotic solution differ from the numerically computed solution, and the uniformly valid solution does fare better. This is given in figure \ref{f:asyimps}. We remark that stronger agreement would be very difficult as the numerics have failed to predict the exact value of $u_0 = 0.5$ at $x=t$. The solutions are computed with $N=200000$!

From these asymptotic predictions, it appears there is a highly oscillatory profile that is $O(1)$ near $x=t \rightarrow \infty$. The fact the growth is not unbounded suggests that even at high mode numbers, the formation of pressure waves may be stable. From the qualitative behaviour of the solution, it makes the highly oscillatory nature of the solutions obtained in section \ref{Imp} less surprising and suggests the numerics are valid in producing such results. It also suggests that the full cylindrical implosion problem becomes increasingly difficult to solve and accurately resolve with increased mode number.

\section{Analysis of Cylindrical System}

\label{Imp}

\subsection{Numerical Approach}

\begin{figure}
 \begin{center}
    \includegraphics[width=5.in]{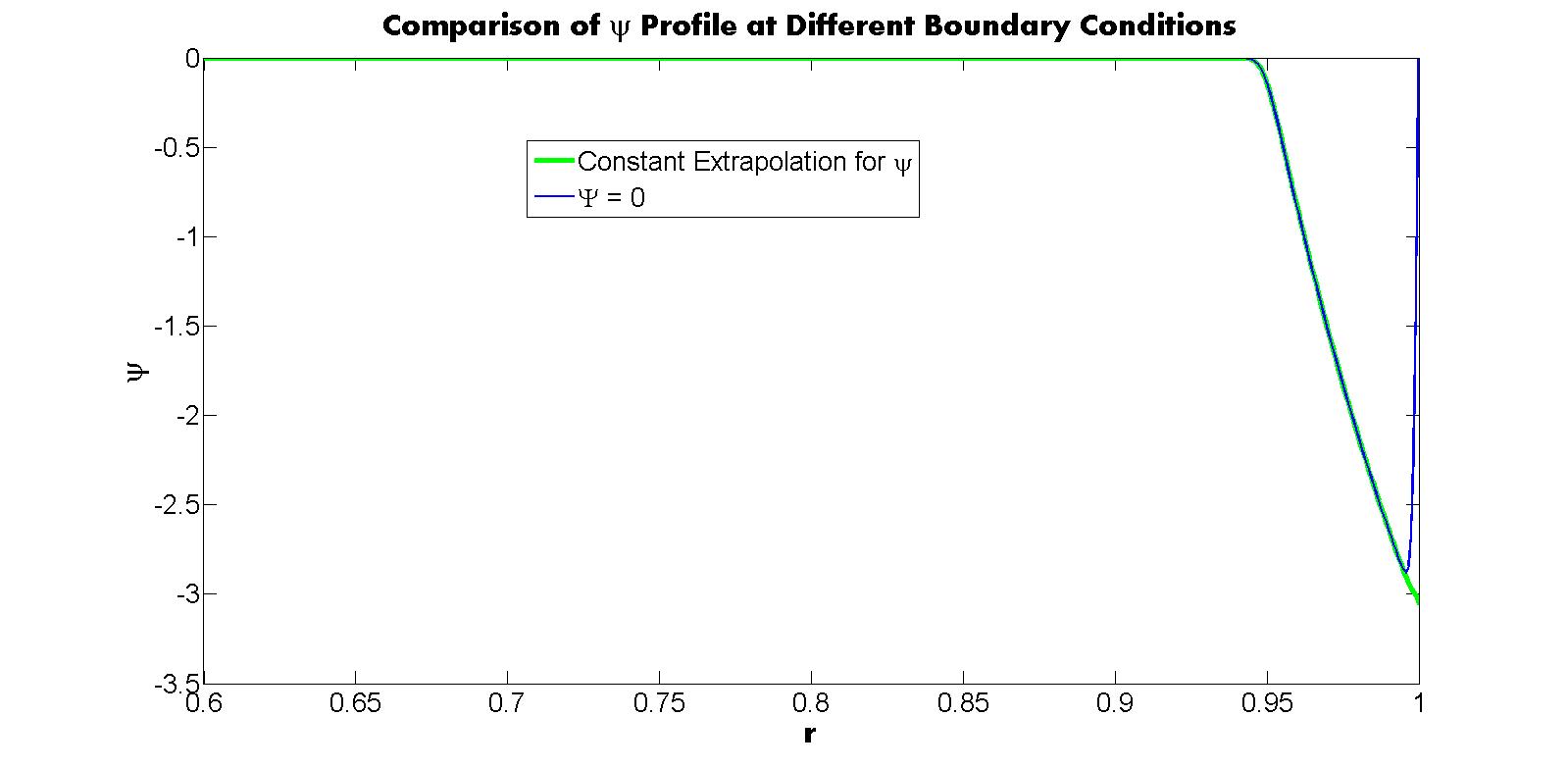}
 \end{center}
\caption{Profile of $\text Im \psi$ at $t=0.01$ and $m=15$ for constant extrapolation and a Dirichlet boundary condition.}
\label{fig:bcq}
\end{figure}

For the case of the systems, we also used a first-order upwinded scheme on a uniform mesh and a split-stepping for the source terms. We follow an analogous approach to section \ref{Burg} although the conservation laws now form a coupled system with various boundary conditions as given by equations (\ref{rho0}) through (\ref{rights}). At the boundaries, when unspecified, we simply use constant extrapolation as this is generally a good approach \cite{Randy}. This can seem unsettling, however this tends to impose absorbing or non-reflecting boundary conditions and maintain consistency with the physical system. From a mathematical point-of-view, the boundary conditions for $\mu_0$ and $s$ being unspecified at $r=1$ stem from the fact that the systems involving these variables have characteristics $v \pm c$, with $v \ll c$ so there are both right and leftgoing characteristics. The leftgoing characteristics emerge from $r=1$ with $\rho_0$ and $\bar{\rho}$ specified, and the rightgoing characteristics are known from the values of $\mu_0$ an $s$ at $t=0$ in combination with knowledge of the value of the densities at $r=1.$ However, the characteristic along which the value of $\psi$ propagates has eigenvalue $v$, which can be of either sign. Using constant extrapolation is well justified when $v>0$ as information is moving to the right. However, when $v<0$, this seems less justified. We try the constant extrapolation at $t=0.01$ in a regime where $v<0$ at $r=1$ and compare that to imposing $\psi=0$ at $r=1$: besides an isolated region at the boundary, the two profiles are identical and we choose constant extrapolation. See figure \ref{fig:bcq}. It appears there is a natural boundary condition that physically emerges by the system itself without specifying one.

\subsection{Numerical Results}

\begin{figure}
 \begin{center}
    \includegraphics[width=2.5in]{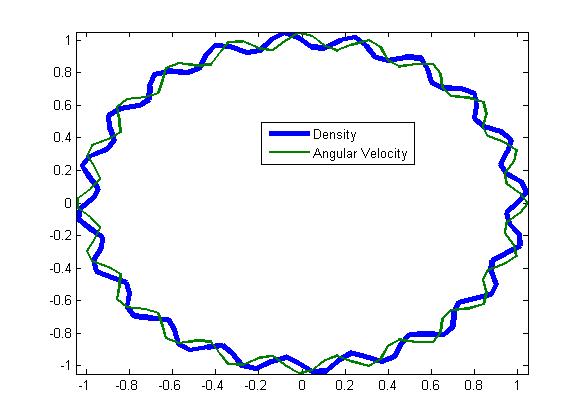}
 \end{center}
\caption{Picture depicting the sinusoidal dependencies of the density and angular velocity.}
\label{f:sinecartoon}
\end{figure}

From the nondimensionalized system, we plot the solution profiles for different values of $m$ at $t=0.05, 0.1,$ and $0.15.$ Some explanation is warranted as to the quantities displayed. As the velocities and densities are real quantities and the system is modelled by a small perturbation modulated by $\e^{i m \theta}$, the values of $\bar{\rho}$ and $s$ are the real parts of the solution as the imaginary parts are zero. The quantity $\psi$ plotted is the imaginary part of the solution $\psi$ as the real part vanishes. This is significant because if the pressure/density perturbations are modulated by cosine terms, the angular velocity is modulated by sine terms. This has a physical significance: the angular velocity should be induced by a pressure/density gradient in the angular direction: with a density modulated by cosine, the angular velocity should be modulated by its derivative, sine. See figure \ref{f:sinecartoon}.

\begin{figure}
 \begin{center}
  $\begin{array}{cc}
    \includegraphics[width=3.5in]{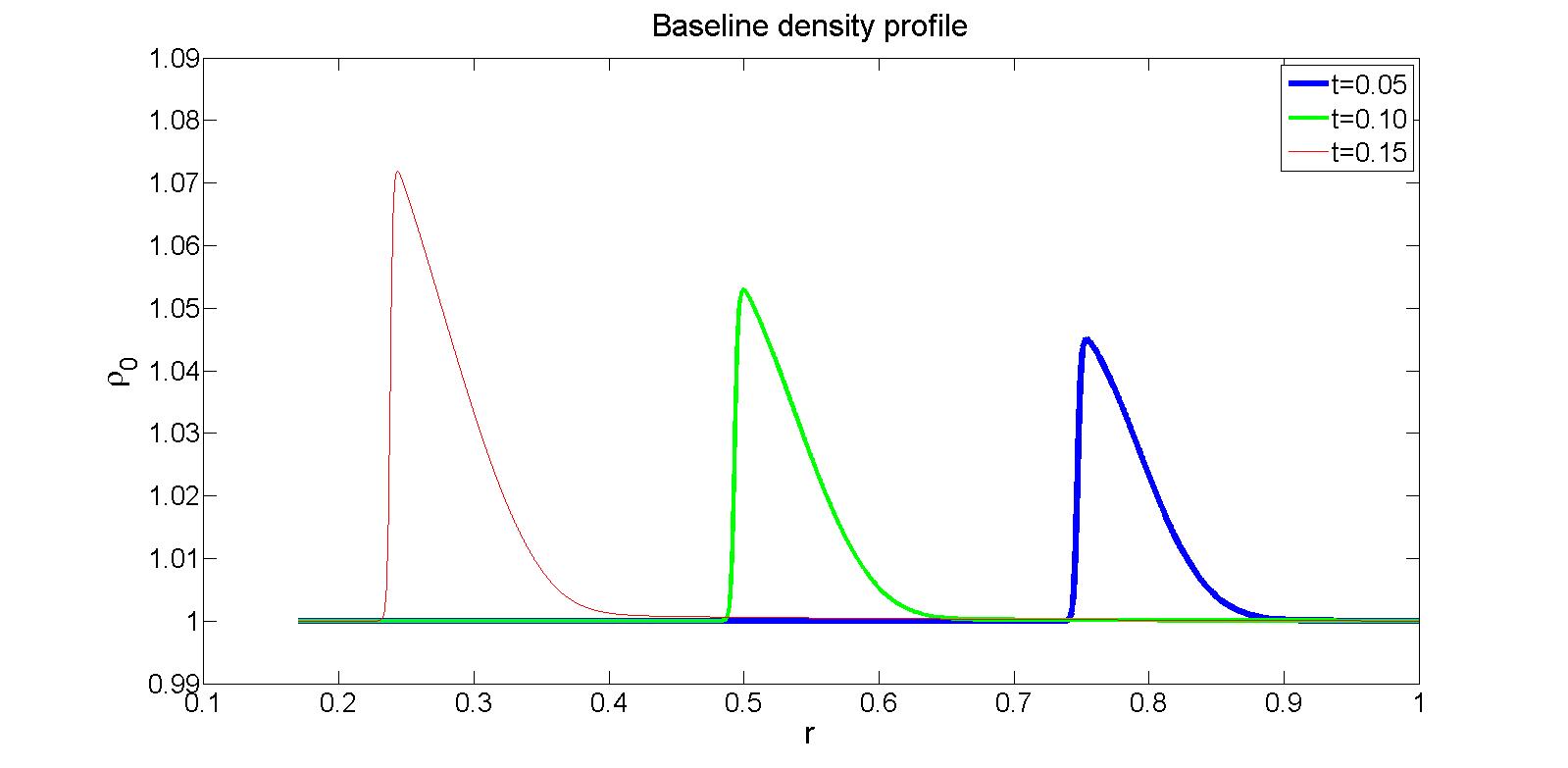} &
    \includegraphics[width=3.5in]{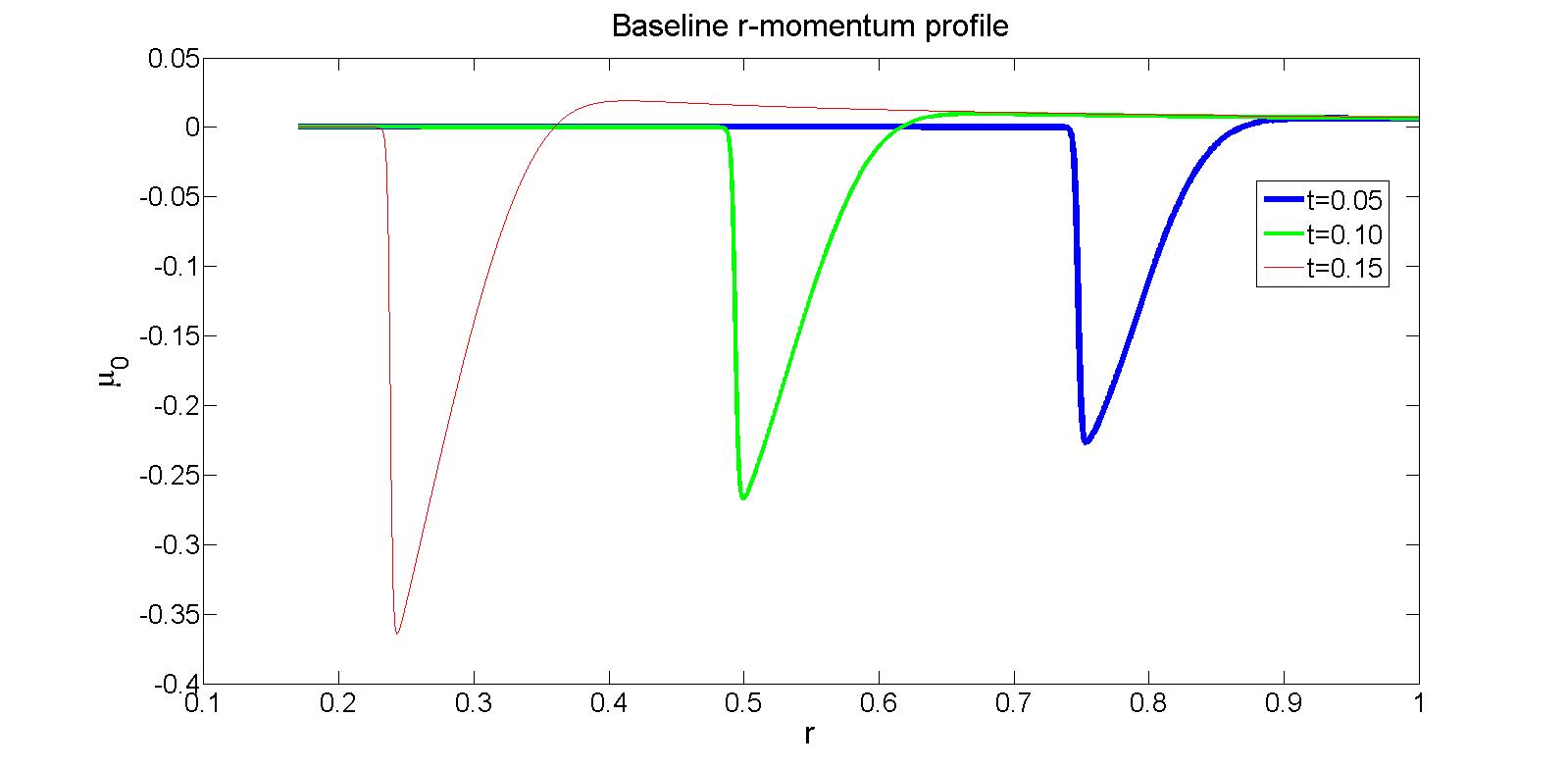}
   \end{array}$
 \end{center}
\caption{The plot of the leading order solutions at $m=0.$ There are no angular perturbations.}
\label{f:m0res}
\end{figure}

\begin{figure}
 \begin{center}
  $\begin{array}{ccc}    
\includegraphics[width=2.35in]{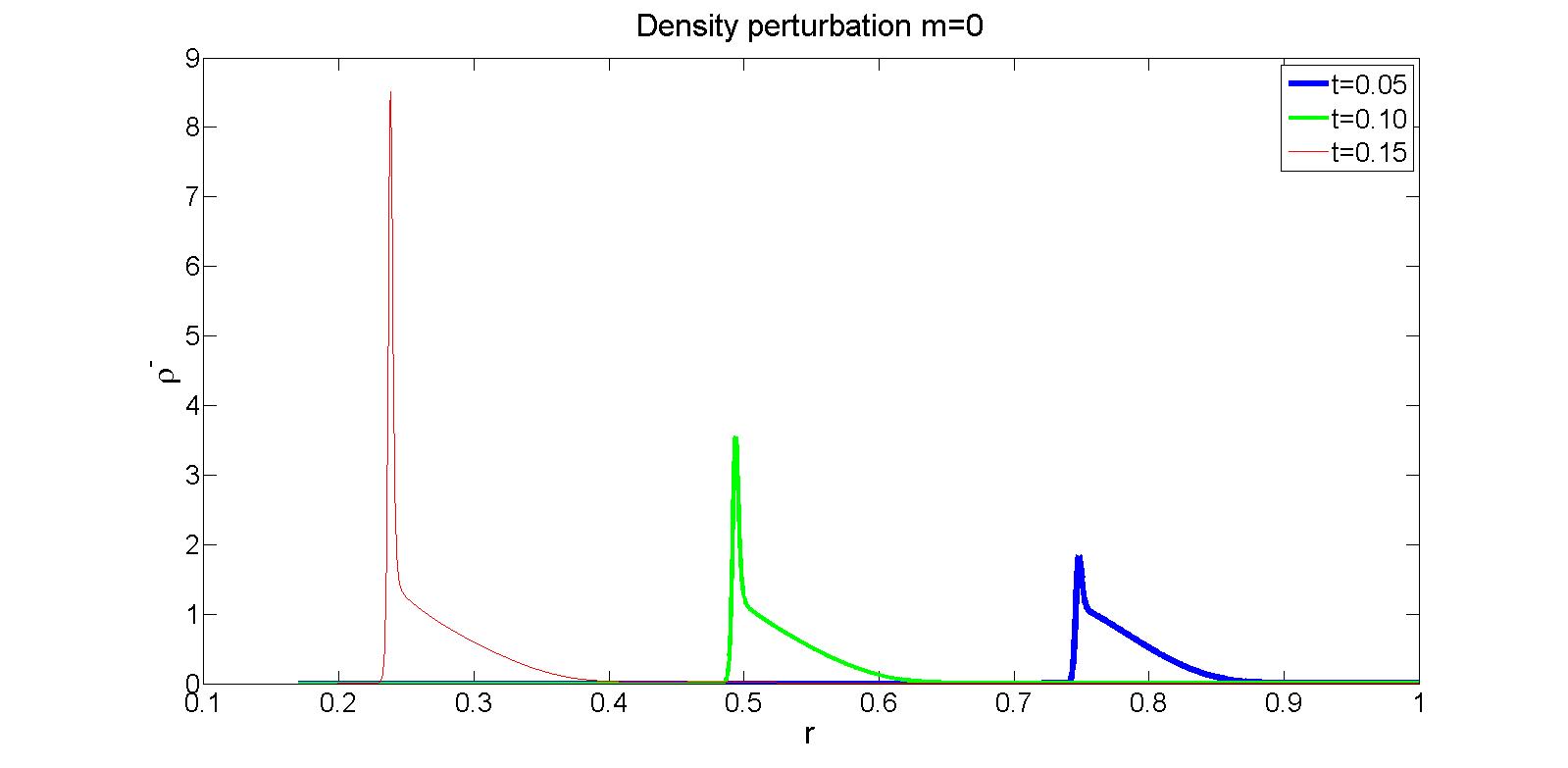} &
\includegraphics[width=2.35in]{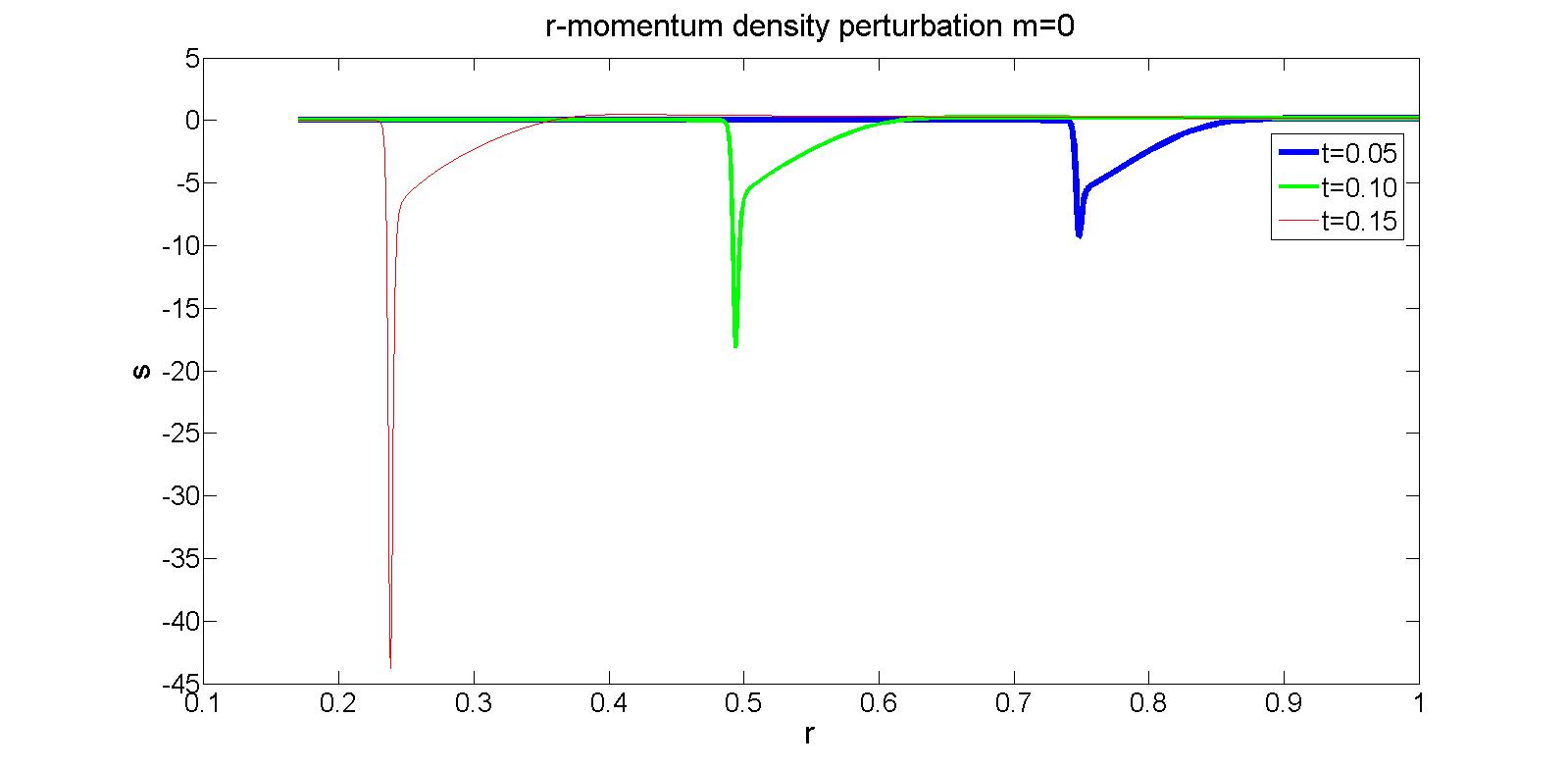} &
\includegraphics[width=2.35in]{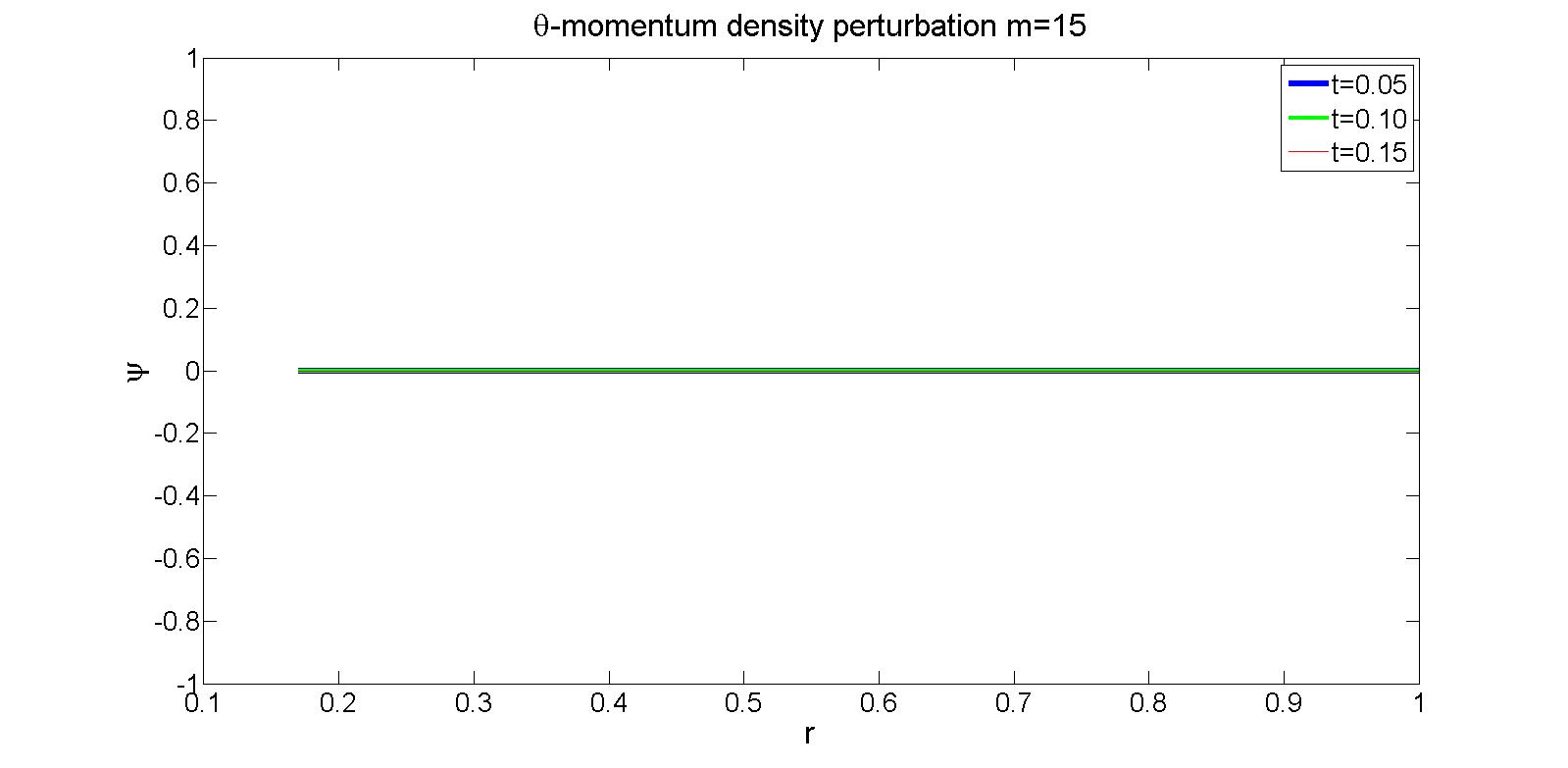} \\
\includegraphics[width=2.35in]{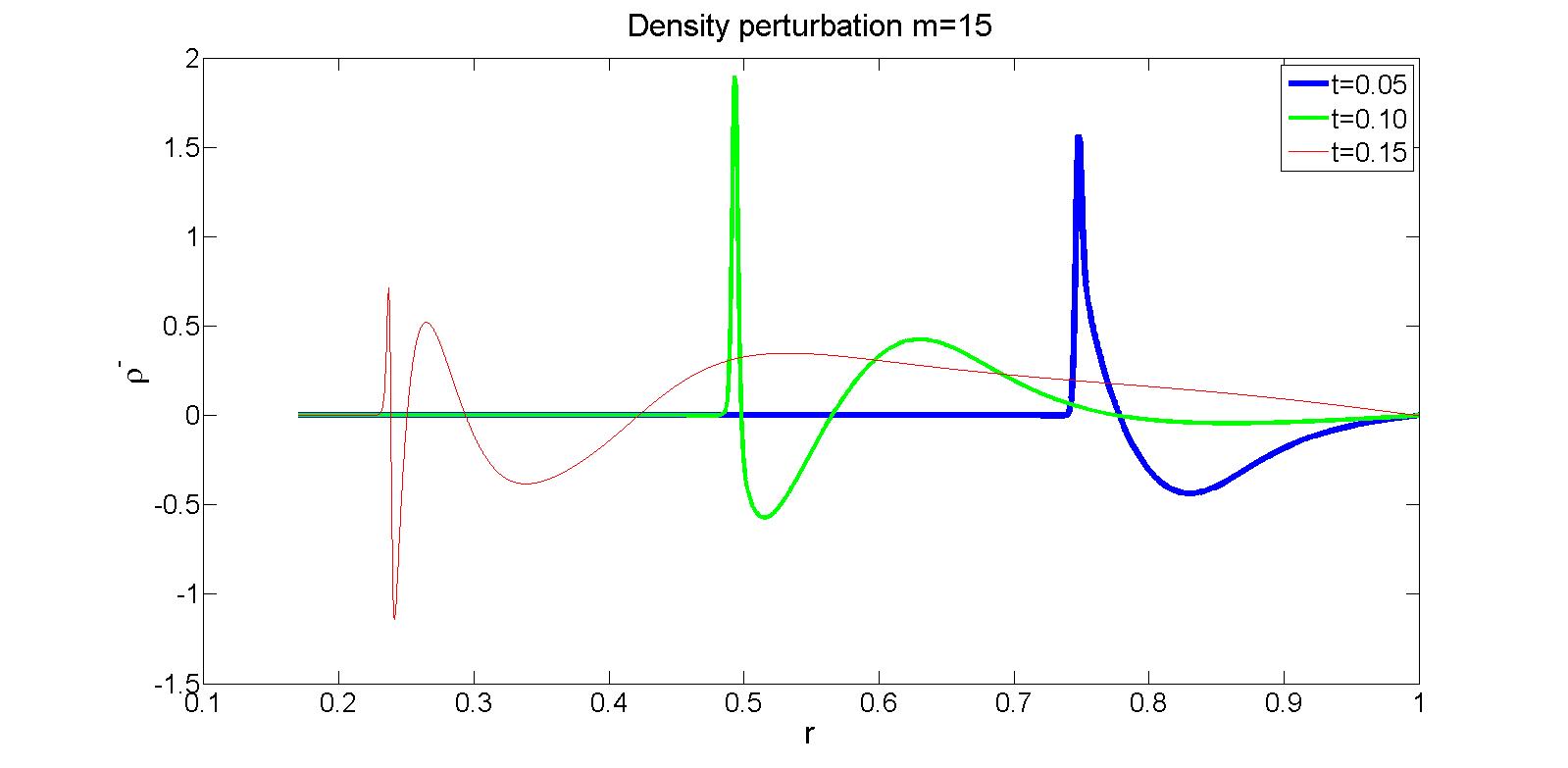} &
\includegraphics[width=2.35in]{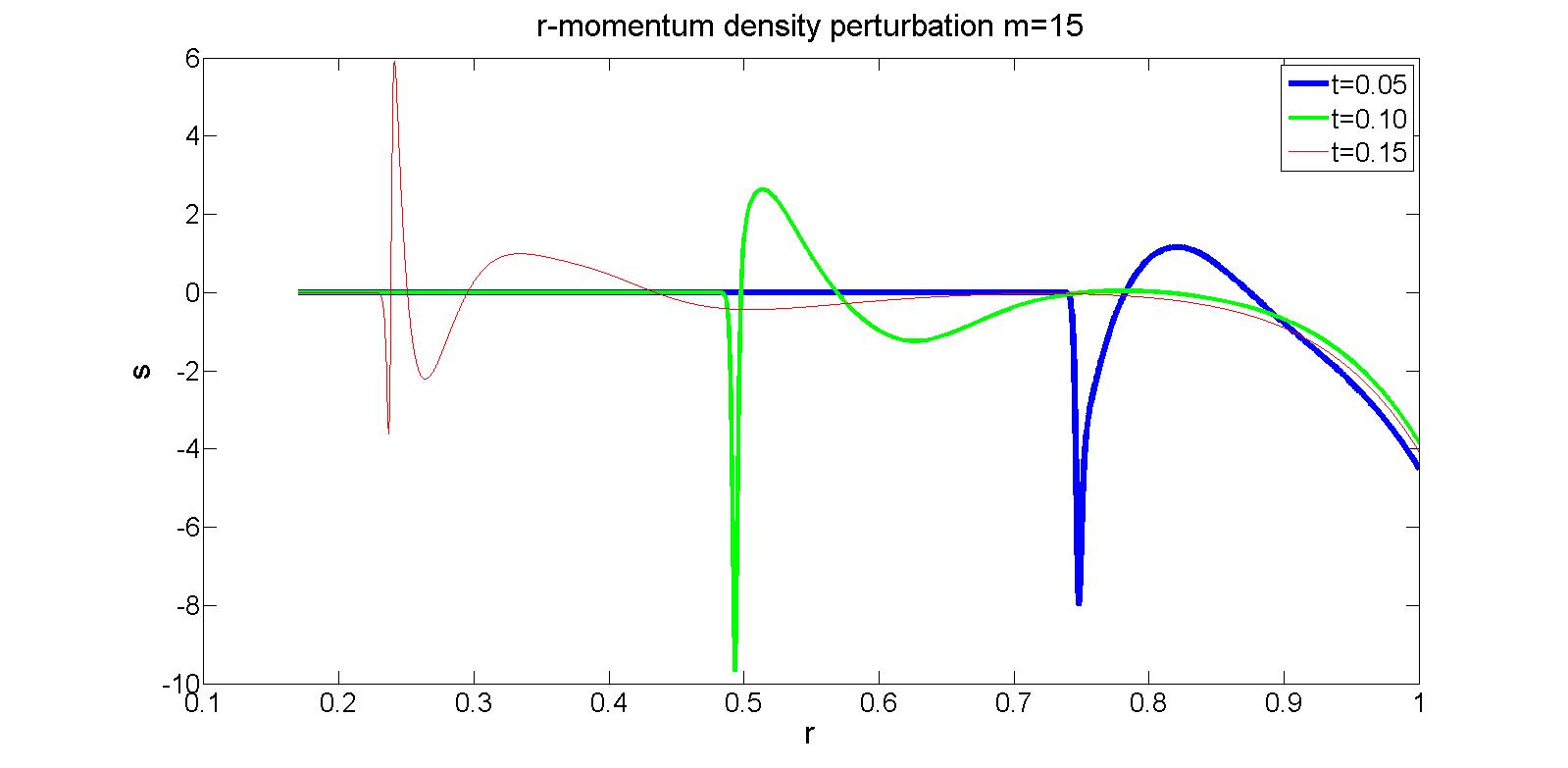} &
\includegraphics[width=2.35in]{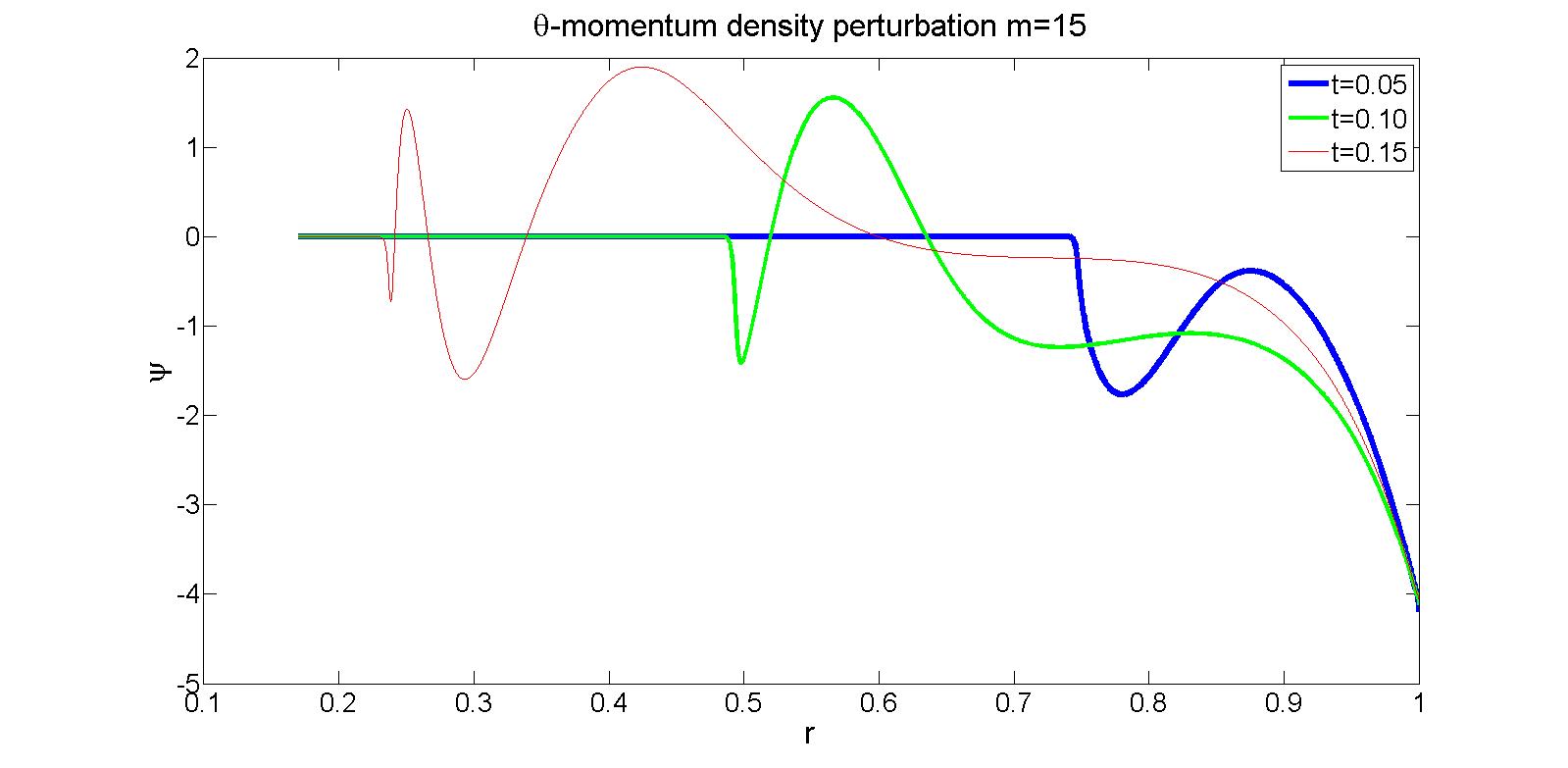} \\
\includegraphics[width=2.35in]{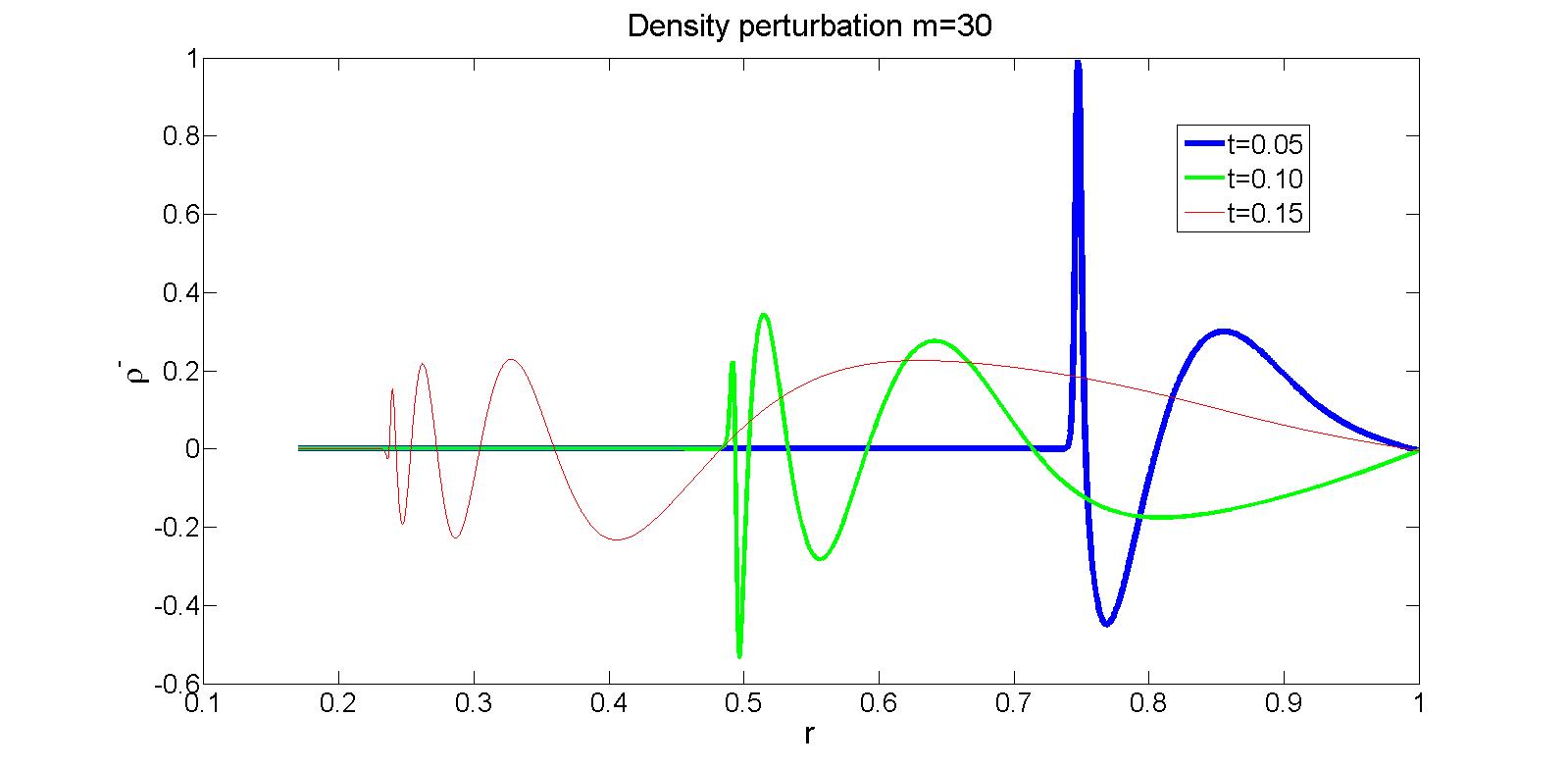} &
\includegraphics[width=2.35in]{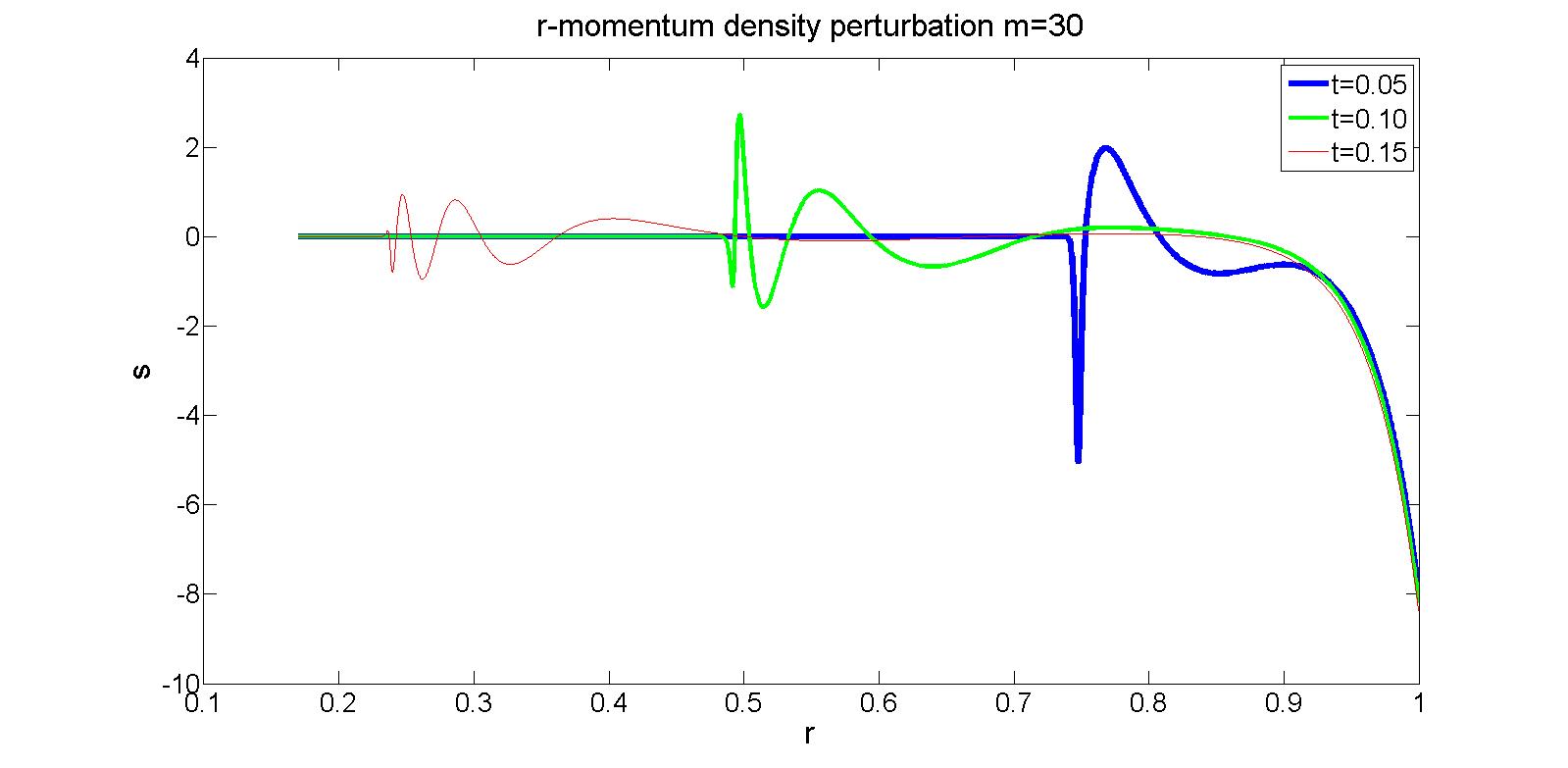} &
\includegraphics[width=2.35in]{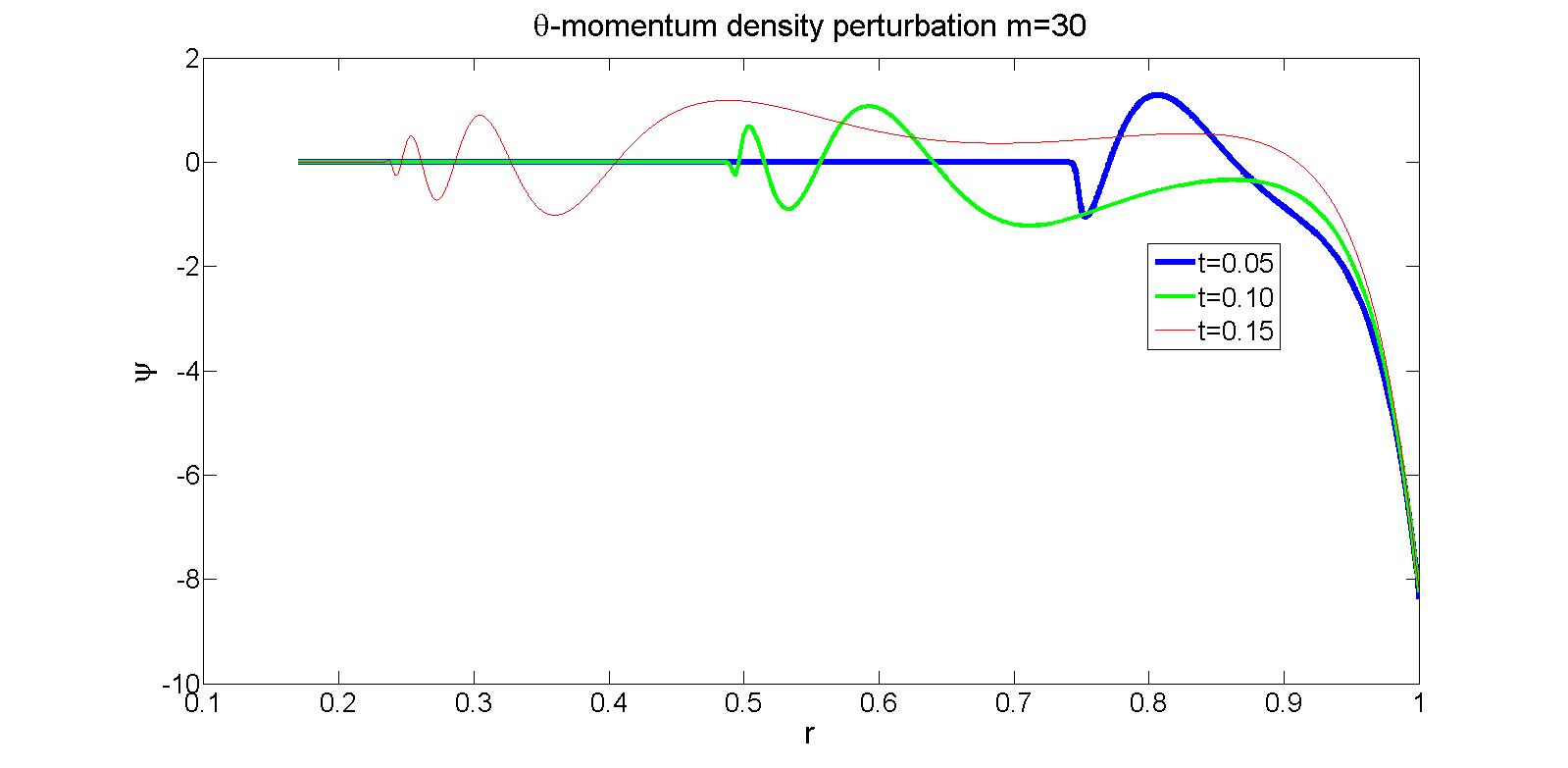} \\
\includegraphics[width=2.35in]{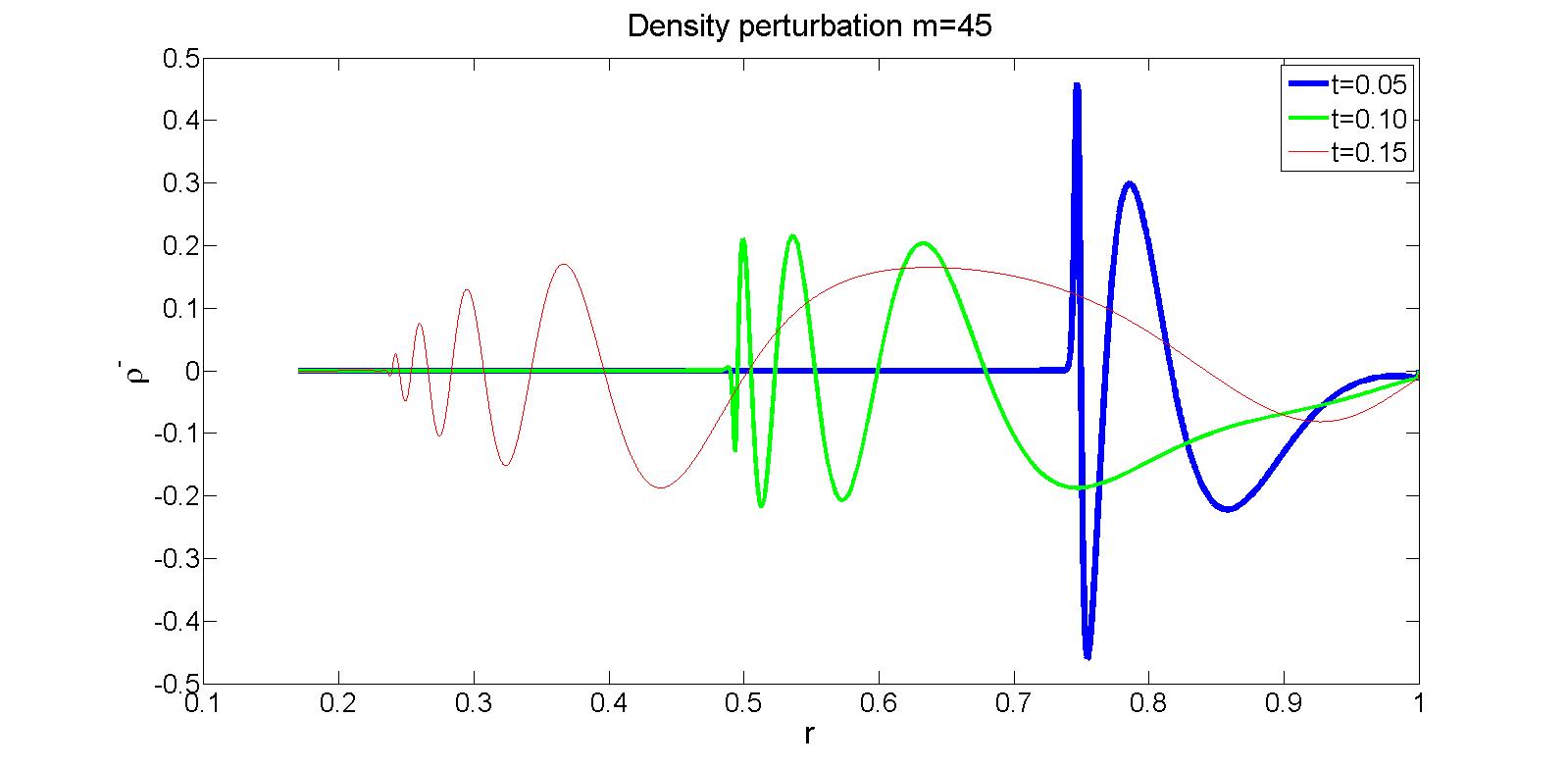} &
\includegraphics[width=2.35in]{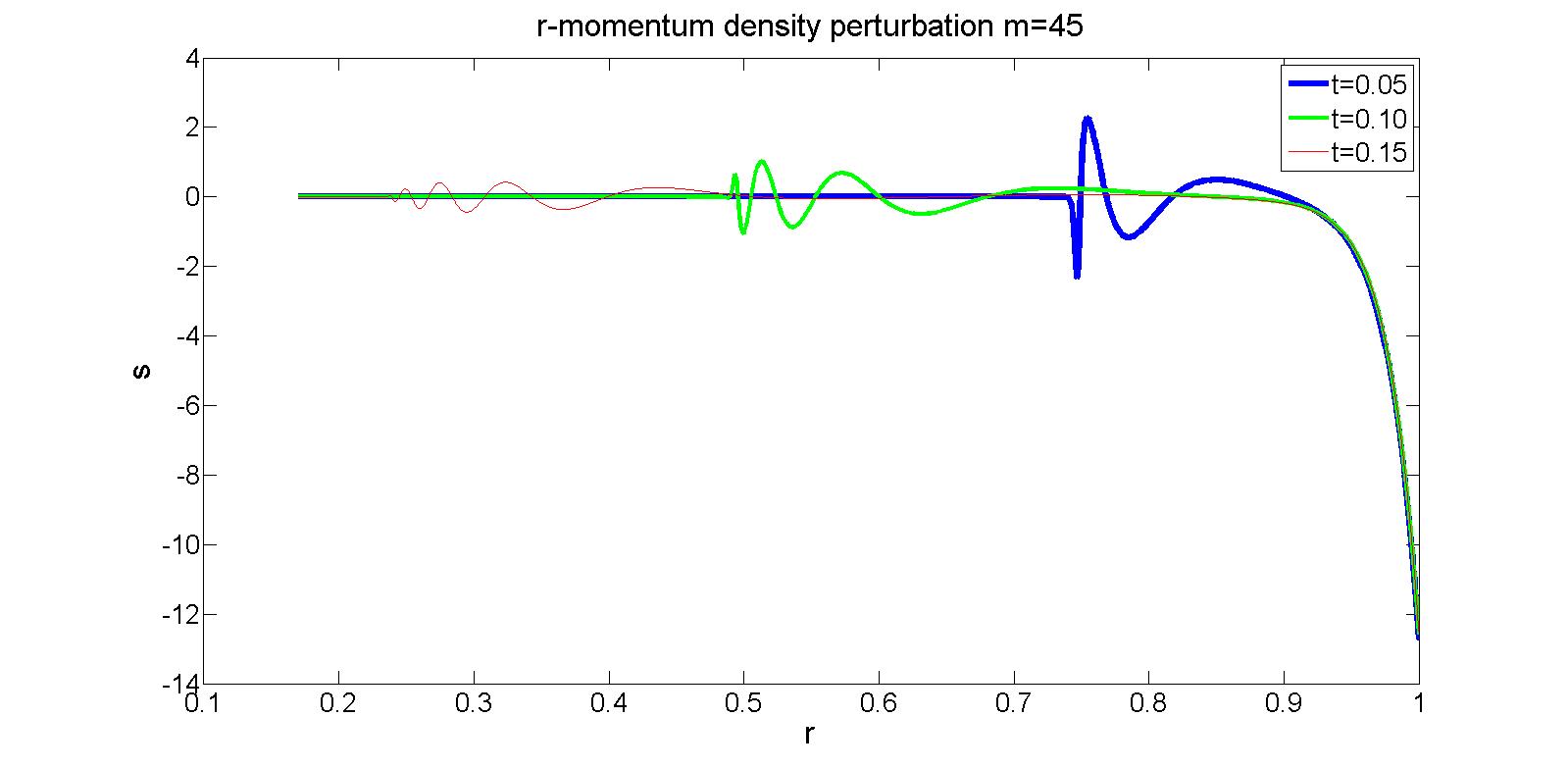} &
\includegraphics[width=2.35in]{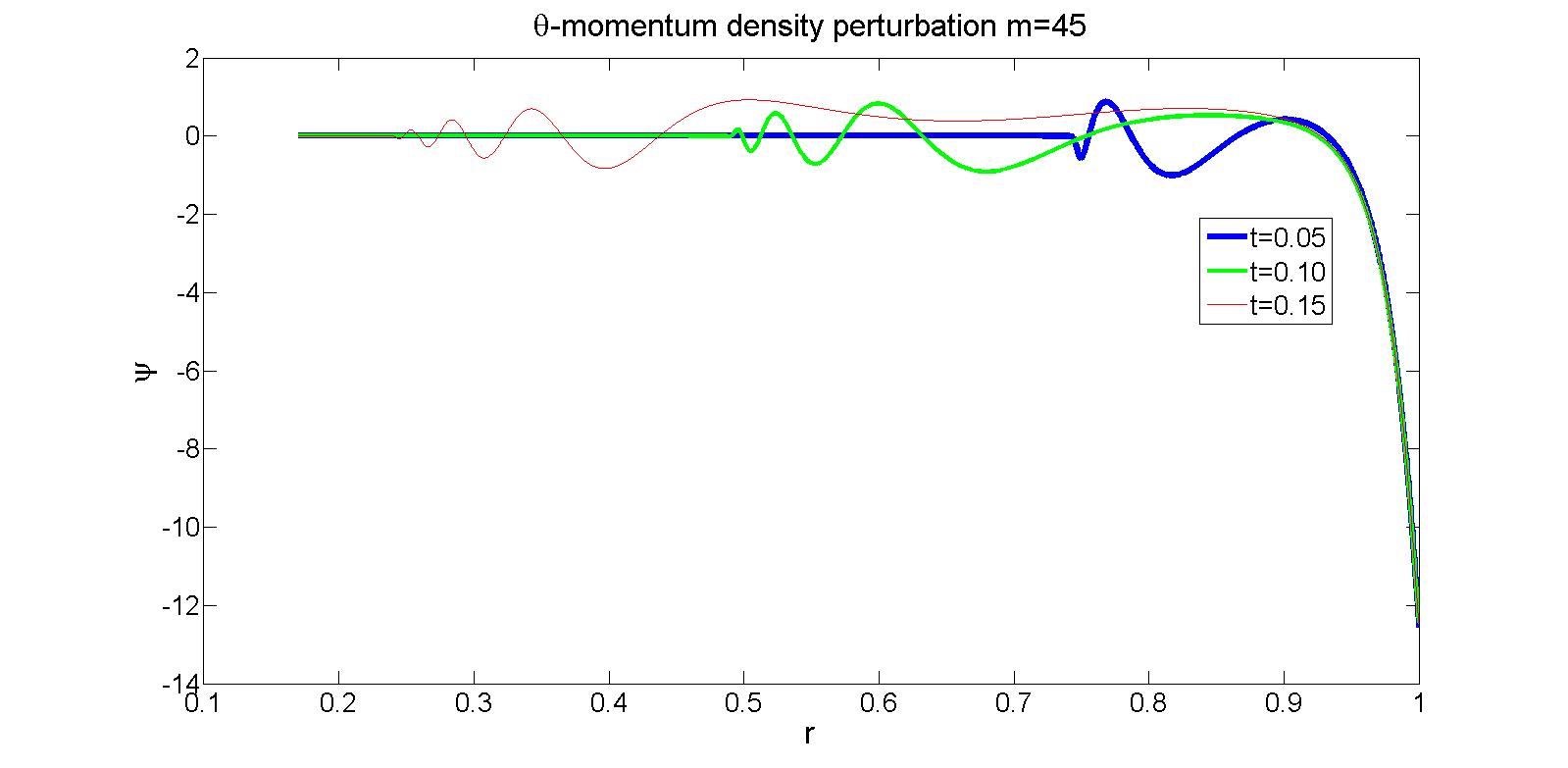} \\
\includegraphics[width=2.35in]{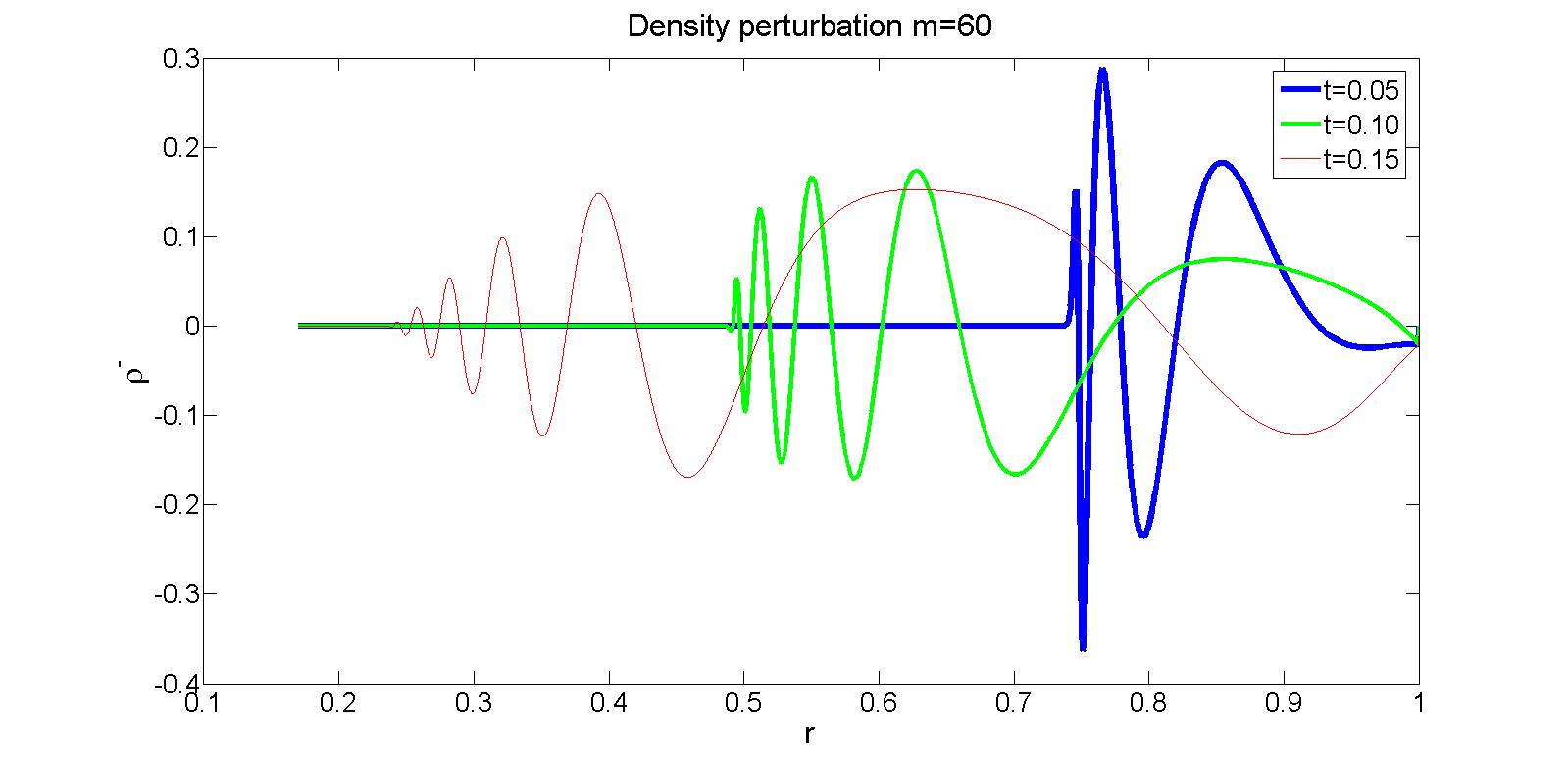} &
\includegraphics[width=2.35in]{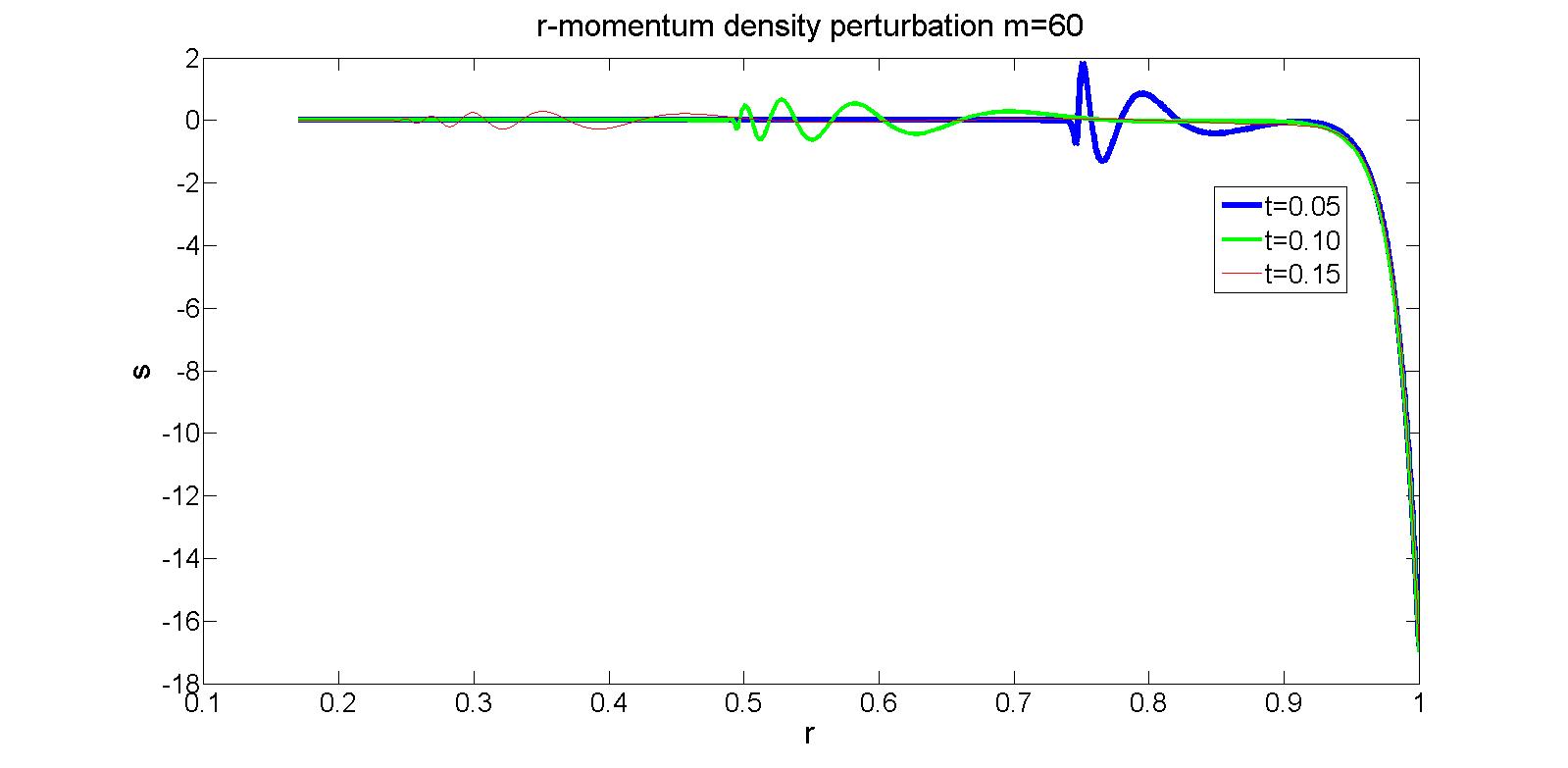} &
\includegraphics[width=2.35in]{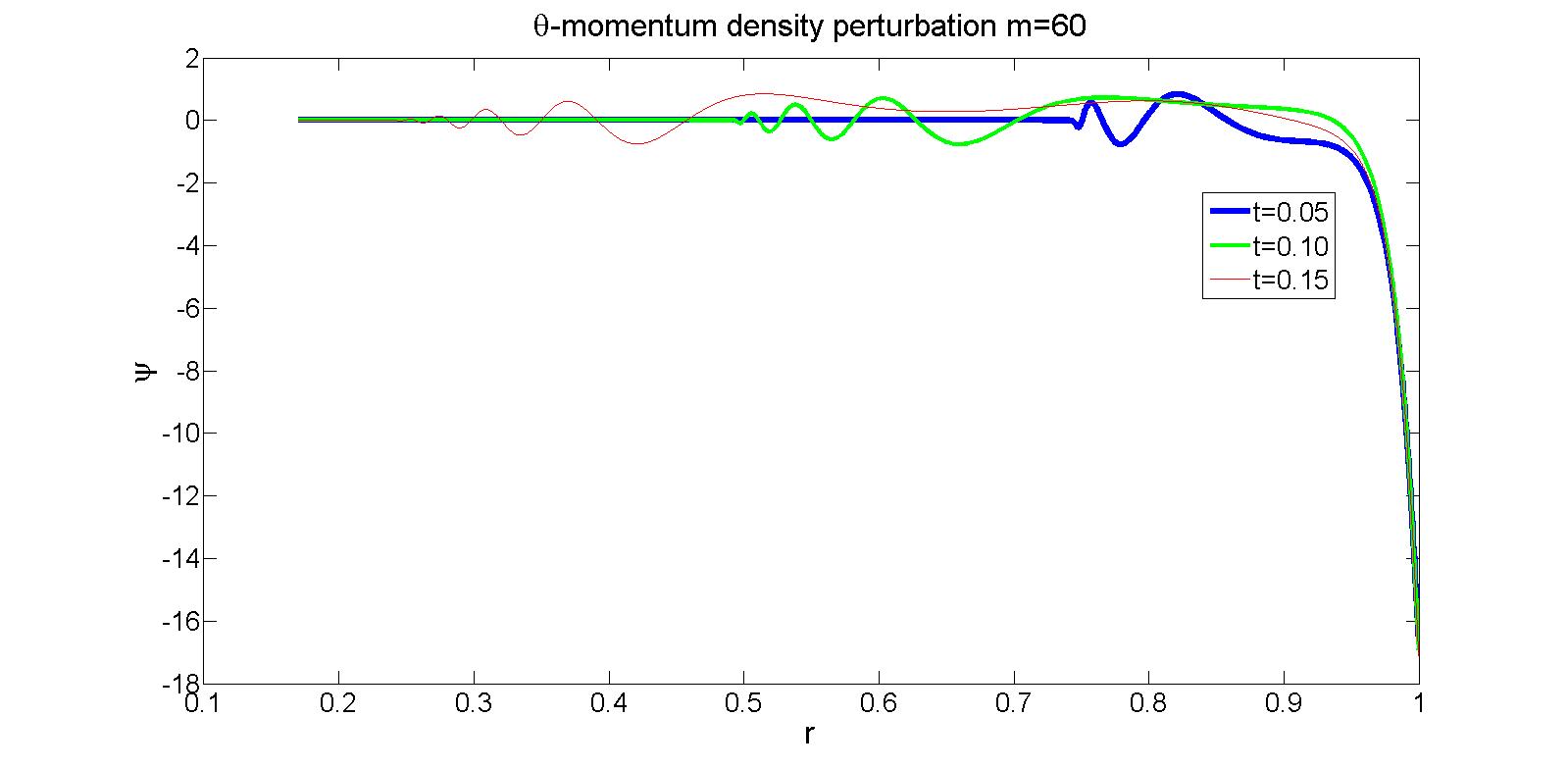}
   \end{array}$
 \end{center}
\caption{Plots of the perturbations at $m=0$, row 1, $m=15$, row 2, $m=30$, row 3, $m=45$, row 4, and $m=60$, row 5. The first column depicts the density perturbation, the second column depicts the $\rhat$-momentum density perturbation, and the third column depicts the $\that$-momentum density perturbation. As expected with $m=0$ there are no angular velocity perturbations.}
\label{f:mresrest}
\end{figure}

We observe that while the disturbance grows for $m=0$, for larger values of $m$, the size of the disturbance shrinks. From the numerical work, we can also estimate the strength of the delta function sources. The work is not entirely trivial as it is difficult to determine based on numerical simulations, when an analytic solution is not known, precisely where a numerical delta function is to be located and what its support is. To estimate the strengths, we obtain two values: first, we compute where the base solutions have reached their peak values (examining numerical plots visually suggests the delta functions have always occurred before this point); and second, because some of the profiles change sign we determine at which point, if any, the perturbed profile changes sign. We then define the rightmost edge of the delta function support as the minimum of these two values (or the point where the base solutions reach their peak values when there is no sign change in the perturbed solution).
As the perturbations are zero to the left of the shock, we numerically integrate up to the cutoff point and define this as our mass of the delta function. Then using linear extrapolation as in section \ref{Burg} with $N=500, ... 4000$, we provide our best possible estimate for the strength of the delta function.

Given the strength of the delta function, we are able to divide this by the jump discontinuity in the base density and base radial momentum density values to estimate the rate of change of the shock front amplitude difference with respect to perturbation amplitude for the $\bar{\rho}$-profile, $\Delta_\eta^{(\bar{\rho})}$, and $\Delta_\eta^{(s)}$ for the $s-$profile, similar to equation (\ref{Delprime}). Physically, it is expected that if the perturbed density front is ahead of or behind the the symmetric front then the perturbed momentum density front should be ahead or behind by the same amount. Indeed, table \ref{t:del} suggests this to be true. We also observe the delta function strengths quickly decay to zero with increasing the mode number. This is perhaps surprising as the physical effects of viscosity have not been included in the model.

The interpretation for the delta function in the angular direction is less clear than above because there is no base solution with an angular velocity. Besides the fact the angular equation does admit a delta function solution, we anticipate its origin due to the presence of delta functions appearing in the density, which effectively induces a pressure gradient in the angular direction localized to a delta function source.

We wish to mention that a value listed as 0.0000 does not mean the numerical solution is identically zero, but to within the precision we are working, its value rounds to zero.

\begin{table}[h]
\begin{center}
\begin{tabular} {c | c | c | c | c | c | c | c | c}
$t$ & m &  $M_{\bar{\rho}}$ & $[\rho_0]$ & $\Delta_\eta^{(\bar{\rho})}$ & $M_s$ & $[\mu_0]$ & $\Delta_\eta^{(s)}$ & $M_\psi$  \\
\hline
\hline
0.05 & 0 & 0.0126 & 0.0450 & -0.2800 & -0.0643 & -0.2267 & -0.2836 & 0 \\
 & 15 & 0.0107 & 0.0450 & -0.2378 & -0.0545 & -0.2267 & -0.2404 & -0.0038 \\
 & 30 & 0.0054 & 0.0450 & -0.1200 & -0.0276 & -0.2267 & -0.1217 & -0.0070 \\
 & 45 & 0.0018 & 0.0450 & -0.0400 & -0.0091 & -0.2267 & -0.0401 & -0.0039 \\
 & 60 & 0.0005 & 0.0450 & -0.0111 & -0.0023 & -0.2267 & -0.0101 & -0.0009 \\
 \hline
 0.10 & 0 & 0.0222 & 0.0529 & -0.4197 & -0.1135 & -0.2670 & -0.4251 & 0  \\
 & 15 & 0.0095 & 0.0529 & -0.1796 & -0.0482 & -0.2670 & -0.1805  & -0.0104 \\
 & 30 & 0.0007 & 0.0529 & -0.0132 & -0.0035 & -0.2670 & -0.0131 & -0.0011 \\
 & 45 & 0.0000 & 0.0529 & 0.0000 & 0.0000 & -0.2670 & 0.0000 & 0.0000 \\
 & 60 & 0.0000 & 0.0529 & 0.0000 & 0.0000 & -0.2670 & 0.0000 & 0.0000 \\
 \hline
 0.15 & 0 & 0.0412 & 0.0718 & -0.5738 & -0.2115 & -0.3641 & -0.5809 & 0 \\
 & 15 & 0.0020 & 0.0718 & -0.0279 & -0.0102 & -0.3641 & -0.0280 & -0.0034 \\
 & 30 & 0.0000 & 0.0718 & 0.0000 & 0.0000 & -0.3641 & 0.0000 & 0.0001 \\
 & 45 & 0.0000 & 0.0718 & 0.0000 & 0.0000 & -0.3641 & 0.0000 & 0.0000 \\
 & 60 & 0.0000 & 0.0718 & 0.0000 & 0.0000 & -0.3641 & 0.0000 & 0.0000 \\
\end{tabular} 
\end{center}
\caption{Tabulation of delta function strengths in perturbations at different times for different azimuthal perturbation modes.} \label{t:del}
\end{table}

\subsection{Delta Function Strengths via ODE}

Following our form in equation (\ref{Mvec}), the strengths of the delta function masses for $\bar{\rho}$ and s, which we denote by $M_{\bar{\rho}}$ and $M_s$ respectively, should obey the system of ordinary differential equations below: 
\begin{align} \frac{\dd}{\dd t} M_{\bar{\rho}} &= \frac{\mu_0^+}{\rho_0^+-1} \bar{\rho}^+ - s^+ - \frac{s^+/r_s + im \psi^+/r_s}{\rho_0^+-1} M_{\bar{\rho}} \label{Mbarrho} \\ \frac{\dd}{\dd t} M_s &= \frac{\mu_0^+}{\rho_0^+-1} s^+ - (c_s^2 \bar{\rho}^+ + 2 \frac{\mu_0^+ s^+}{\rho_0^+} - \frac{(\mu_0^+)^2 \bar{\rho}^+}{\rho_0^+}) - ( \frac{1}{r_s} \frac{2 \mu_0^+ s^+}{\rho_0^+} + \frac{1}{r_s} \frac{(\mu_0^+)^2 \bar{\rho}^+}{(\rho_0^+)^2} - \frac{i m}{r_s} \frac{\mu_0^+ \psi^+}{\rho_0^+} ) M_{s} \label{Ms} 
\end{align}
 where $r_s$ is the shock position. In obtaining these equations we used the fact that to the left of the shock, only $\rho_0 = 1$ is nonzero.

Verifying this numerically by a direct algorithm is difficult as defining precisely the value of a function on either side of a shock can scarcely be done by visual inspection for solutions of the profiles obtained, let alone defining a generic algorithm that always defines the values on either side of the shock correctly. Indeed, some profiles do not change their sign whereas others do. This is compounded by our observation that the solution tends to oscillate more quickly with larger mode numbers and that to within the numerical precision used, it appears from table \ref{t:del} the delta function masses may only be accurate to within around $0.001$.

To obtain evidence for equations (\ref{Mbarrho}) and (\ref{Ms}), we choose a fixed time of $t=0.10$ and carefully study the profiles with $m=30$. By combining visual inspection and numerical searches for peak values, we estimate: \\ $\rho_0^+ \approx 1.0529, \mu_0^+ \approx -0.2670, \bar{\rho}^+ \approx -0.5326, s^+ \approx 2.7320, \psi^+ \approx 0.1022, r_s \approx 0.4997, M_{\rho} \approx 0.000798, M_s \approx -0.004023. $ 
We then step discretely forward in time with a time step of $\Delta t = 0.0003$ using the derivatives at $t=0.10$, from (\ref{Mvec}), and compare against the numerical values. The results can be seen in table \ref{tab:weaksupport}. The ODE-prediction appears sensible, agreeing in whether the delta-mass increases/decreases, although this analysis is very loose. As the jump in the momentum in the angular direction is zero in the $O(1)$ system, we do not formulate an ODE for $M_{\phi}$, although one could be obtained by replacing $s_1(t)$ in (\ref{matMass}) in terms of a different delta function mass.

\begin{table}[h]
\begin{center}
\begin{tabular} {c | c | c | c }
Quantity & Initial Value & ODE Prediction & Numerical Value \\
\hline
$M_{\bar{\rho}}$ & 0.0007975 & 0.0007874 & 0.0007670 \\
$M_s$ & -0.004023 & -0.003836 & -0.0038501 \\
\end{tabular} 
\end{center}
\caption{ODE prediction and numerical prediction for delta function masses at $t=0.1003$.} \label{tab:weaksupport}
\end{table}

\section{Summary and Future Outlook}

\label{Con}

This work suggests that having on the order of 100 pistons in the MTF nuclear reactor design of General Fusion \cite{GF} is a promising strategy for reducing unwanted effects of asymmetries: perturbations induced by low azimuthal mode numbers, besides $m=0$ which do not contribute to asymmetric distortions, do propagate but the higher mode numbers tend to dissipate through the focusing thereby extinguishing their effects. The work presented in \cite{GF-RM} suggests that the high mode numbers are more problematic in the context of nuclear fusion. The study just cited explored the interaction of the molten metal and the plasma with azimuthal perturbations beginning when the molten metal and plasma interact; it was found that high azimuthal mode numbers cause the greatest problems for plasma-metal interactions, resulting in mixing, and low mode numbers pose little problem. From this we can infer, based on this current work, a positive outlook for the reactor design. 


We would like to acknowledge a number of limitations to the current model, which should be considered carefully in interpreting these results. This work is based on a linear perturbation; this is a great idealization of an infinitely tiny asymmetry. Our model did not take into account fluid viscosity or the rotation of the molten metal. The equation of state was taken to be linear but more sophisticated equations of state could be used. Furthermore, a cylinder is an idealization of the actual geometry, which is far more complex, and would require intense study to fully model and understand.

Our work included some novel steps that develop mathematical and physical understanding of implosion processes including the need to solve nonhomogeneous conservation laws with singular sources, and the acquisition of an asymptotic solution to the linear Klein-Gordon equation. We have also found some interesting open problems such as how to better measure the delta-function strength in these more complex systems of equations, how the results would vary in the case of a sphere where the first order system of conservation laws would not be directly diagonalizable, and whether numerical viscosity effects could be the cause of the diminishing delta-strengths. There are also some possibly theoretical open problems in the numerical treatment of the delta functions and how the generalized tangent vectors lead to the measure-valued solutions when given various linearizations.

\section*{Acknowledgments}

The author would like to acknowledge a number of individuals for valuable discussions and ideas in completing this work. A thanks to John Garnett for suggesting the convergent contour integral with $x>t$; to James Ralston for a fruitful discussion on the Bateman manuscript integrals and qualitative behaviour of such integrals; to Brian Wetton for suggesting studying the shock front difference and providing commentary; to Russ Caflisch for a discussion of the underlying physics; and to Andrea Bertozzi for feedback on the scope of the work. The author also appreciates the comments of the reviewer to add a more theoretical discussion of the system.

\end{document}